\documentclass[12pt,twocolumn]{openjournal}

\usepackage{lipsum}

\usepackage{xcolor}
\usepackage{textgreek}
\usepackage[utf8]{inputenc}
\usepackage[english]{babel}
\usepackage{hyperref}
\hypersetup{
    colorlinks=true,
    linkcolor=blue,
    filecolor=blue,      
    urlcolor=red,
    citecolor=blue,
}
\usepackage{color,colortbl}
\usepackage{tensind}
\tensordelimiter{?}
\DeclareGraphicsExtensions{.bmp,.png,.jpg,.pdf}
\usepackage{verbatim}
\usepackage[normalem]{ulem}
\usepackage{orcidlink}
\usepackage{soul}

\graphicspath{ {./figs/} }

\begin{document}
\title{The cosmological population of Gamma-ray Bursts from the disks of Active Galactic Nuclei}

\shorttitle{GRBs in AGN disks}

\author{Hoyoung D. Kang$^1$}
\author{Rosalba Perna$^1$}
\author{Davide Lazzati$^2$}
\author{Yi-Han Wang$^3$}

\affiliation{$^1$Department of Physics and Astronomy, Stony Brook University, Stony Brook, NY 11794-3800, USA}
\affiliation{$^2$Department of Physics, Oregon State University, 301 Weniger Hall, Corvallis, OR 97331, USA}
\affiliation{$^3$Nevada Center for Astrophysics, University of Nevada, 4505 S. Maryland Pkwy., Las Vegas, 89154, NV, USA}

\shortauthors{H. D. Kang et al.}

\begin{abstract}
With the discovery of gravitational waves (GWs), the disks of Active Galactic Nuclei (AGN) have emerged as 
{an}
interesting environment for hosting  a fraction of their
sources. AGN disks are conducive to forming both long and short Gamma-Ray Bursts (GRBs), and their anticipated cosmological occurrence within these disks has potential to serve as an independent tool for probing and calibrating the population of stars and compact objects within them, and their contribution to the GW-detected population. In this study, we employ Monte Carlo methods in conjunction with models for GRB electromagnetic emission in extremely dense media to simulate the cosmological occurrence of both long and short GRBs within AGN disks, while also estimating their detectability across a range of wavelengths, from gamma-rays to radio frequencies. {We investigate two extreme scenarios: ``undiffused", in which the radiation escapes without significant scattering (i.e. if the progenitor has excavated a funnel within the disk), and ``diffused", in which the radiation is propagated through the high-density medium, potentially scattered and absorbed. In the diffused case, we find that 
 the majority of detectable GRBs, which are at most a few percent of the total,
 are likely to originate from relatively low redshifts, and from the outermost regions of large  supermassive black hole (SMBH) masses,  $\gtrsim 10^{7.5} \rm M_{\odot}$. In the undiffused case, which has a GRB detection probability $\sim 40-50\%$, we expect a similar trend, but with a considerable contribution from the intermediate regions of lower SMBH masses. Detectable emission is generally expected to be dominant in prompt $\gamma$-rays if diffusion is not dominant, and X-ray afterglow if diffusion is important; however, the nature of the dominant observable signal highly depends on the specific AGN disk model}, hence making GRBs in AGN disks also potential probes of the structure of the disks themselves.
\end{abstract}


\maketitle

\section{Introduction}
\label{sec:intro}

The study of Gamma-Ray Bursts (GRBs), with their long and short classes, has a history spanning more than three decades. 
Traditionally, these sources have been linked to galactic environments characterized by relatively modest ambient densities, $n$,  ranging from a few to hundreds of atoms per cm$^{3}$ \citep{Scalo2001, Holland2002, Muccino2013, Krongold2013, Fong2015}.

In recent years, however, there has been a growing interest in the possibility that some of these GRBs might originate within the disks of Active Galactic Nuclei (AGNs), where the densities are substantially higher. This idea has been partly motivated by the potential of AGN disks to explain certain unexpected observations in gravitational wave data, including the detection of black holes (BHs) across the low \citep{Abbott2020a} and high mass \citep{Abbott2020prl} gap, eccentric mergers \citep{Samsing2022},
as well as an asymmetry in the distribution of the spins  of BHs produced in mergers \citep{Callister2021, McKernan2022chi,Wang2021chi}. 

The fact that both Long GRBs (LGRBs) and Short GRBs (SGRBs) could occur in AGN disks is not surprising. Stars are believed to exist within AGN disks, either forming in situ due to gravitational instabilities in the outer disk (e.g. \citealt{Goodman2003, Chen2023}) 
or due to capture from the nuclear star cluster surrounding the AGN (e.g. \citealt{Artymowicz1993, Fabj2020,Wang2024}). The evolution of stars in these dense environments has been extensively studied recently \citep{Cantiello2021,Jermyn2021, Dittman2023, Chen2024, Liu2024}. These studies have revealed that AGN stars, prior to going supernova, tend to grow to substantial masses and are characterized by high rotational speeds (but see \citealt{Chen2023b}, if strong disk turbulence is taken into account), making them ideal candidates as progenitors of LGRBs. Additionally, AGN disks are conducive to the formation of SGRBs, arising from binary mergers of neutron stars (NS) and NS-BH mergers with suitably small mass ratios. This propensity for compact object binary formation in AGN disks is due to the presence of migration traps, facilitating binary formation through dynamical interactions \citep{Bellovary16,
Tagawa2020, Perna2021b, Grishin2024}. Furthermore, the interaction of these objects with the dense AGN disk medium leads to the loss of kinetic energy, further promoting binary formation\citep{Rowan2023,Li2023}.

Recently a candidate for a GRB in a disk has been proposed (\citealt{Levan2023,Lazzati2023}, but see \citealt{Stratta2024}), lending credibility to these theoretical predictions\footnote{Note also the case of GRB~150101B hosted in a galaxy with an AGN \citep{Xie2016}. However, its offset from the centre does not make this GRB a candidate for an AGN-disk host.}. However, there has been no comprehensive analysis connecting the underlying AGN stellar population to the observable occurrence of both LGRBs and SGRBs. This paper aims to address this gap.
Specifically, we  employ Monte Carlo statistical techniques, coupled with models for the GRB electromagnetic emissions in highly dense media, to predict the properties of the population of observable GRBs originating from AGN disks. 
Even though this population is expected
to be largely subdominant with respect to the global observed GRB population,
by providing theoretical predictions of its prompt and afterglow emission which can be compared with observations, we hope to:
({\em a}) Impose constraints on the stellar and compact object populations within AGN disks, and thus in turn help calibrate 
 the proportion of binary BH merger events attributed to the AGN channel;
 ({\em b}) Quantify the contribution of GRB sources to the unknown variability observed in AGNs; ({\em c}) Identify features which are sensitive to the disk structure, and which can in turn help probe the structure of the AGN disk itself. 

\begin{figure}
    \centering
\includegraphics[width=0.45\textwidth]{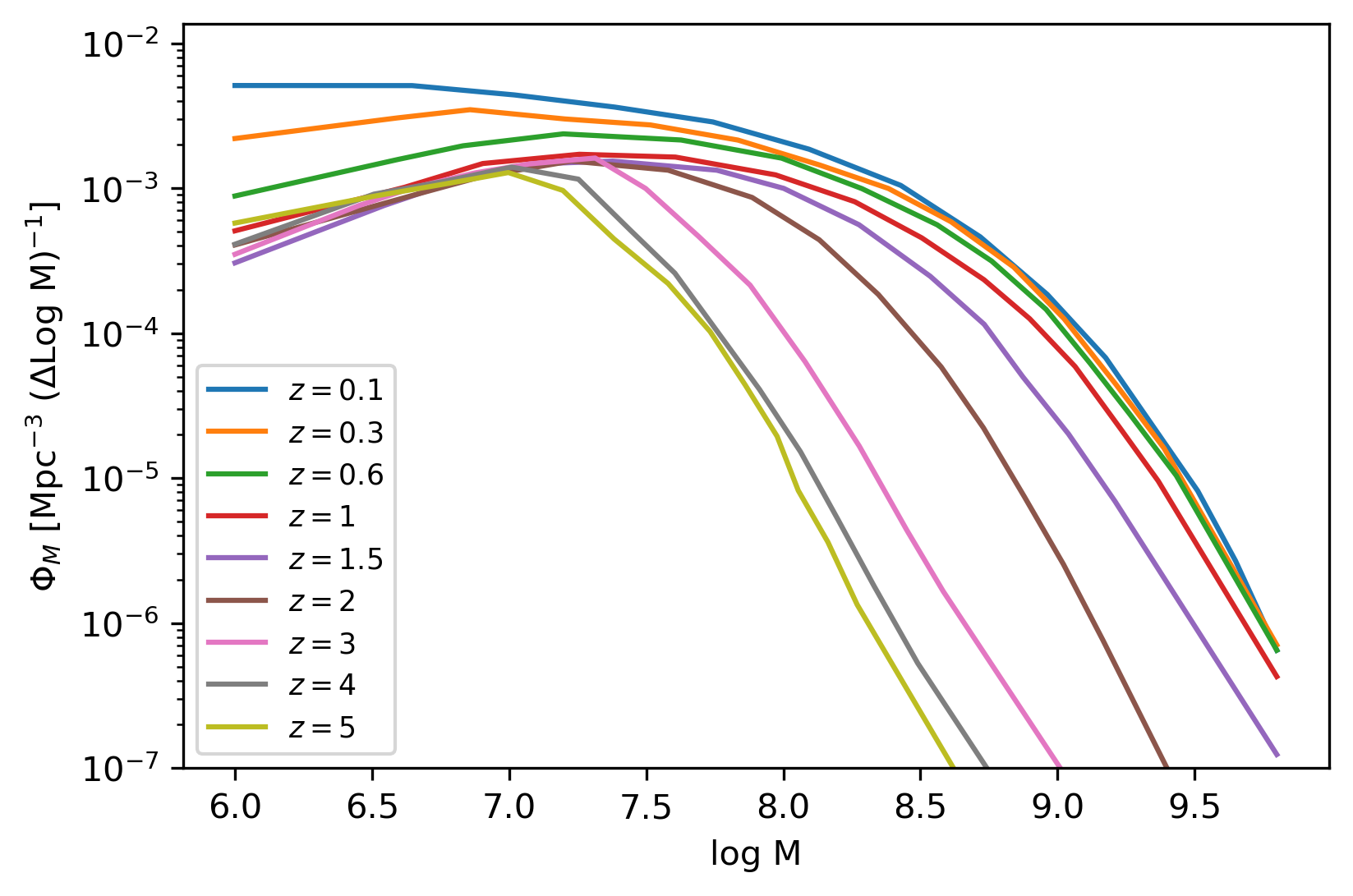}      
    \caption{\small {Redshift evolution of the global SMBH mass function density. The mass profile has been reconstructed by interpolating the mass functions from \citet{Merloni2008}.   } }
    \label{fig:BHMF}
\end{figure}
\vspace{0.1in}

The paper is organized as follows. Sec.~2 describes the methods and the underlying inputs of the analysis, from the 
cosmological supermassive black hole (SMBH) distribution, to the disk models, to the GRB light curves in very dense media. 
Results from the Monte Carlo simulation of the cosmological GRB population  are presented in Sec.~3.  We summarize and conclude
in Sec.~4.

\begin{figure*}
    \centering
    \includegraphics[width=0.45\textwidth]{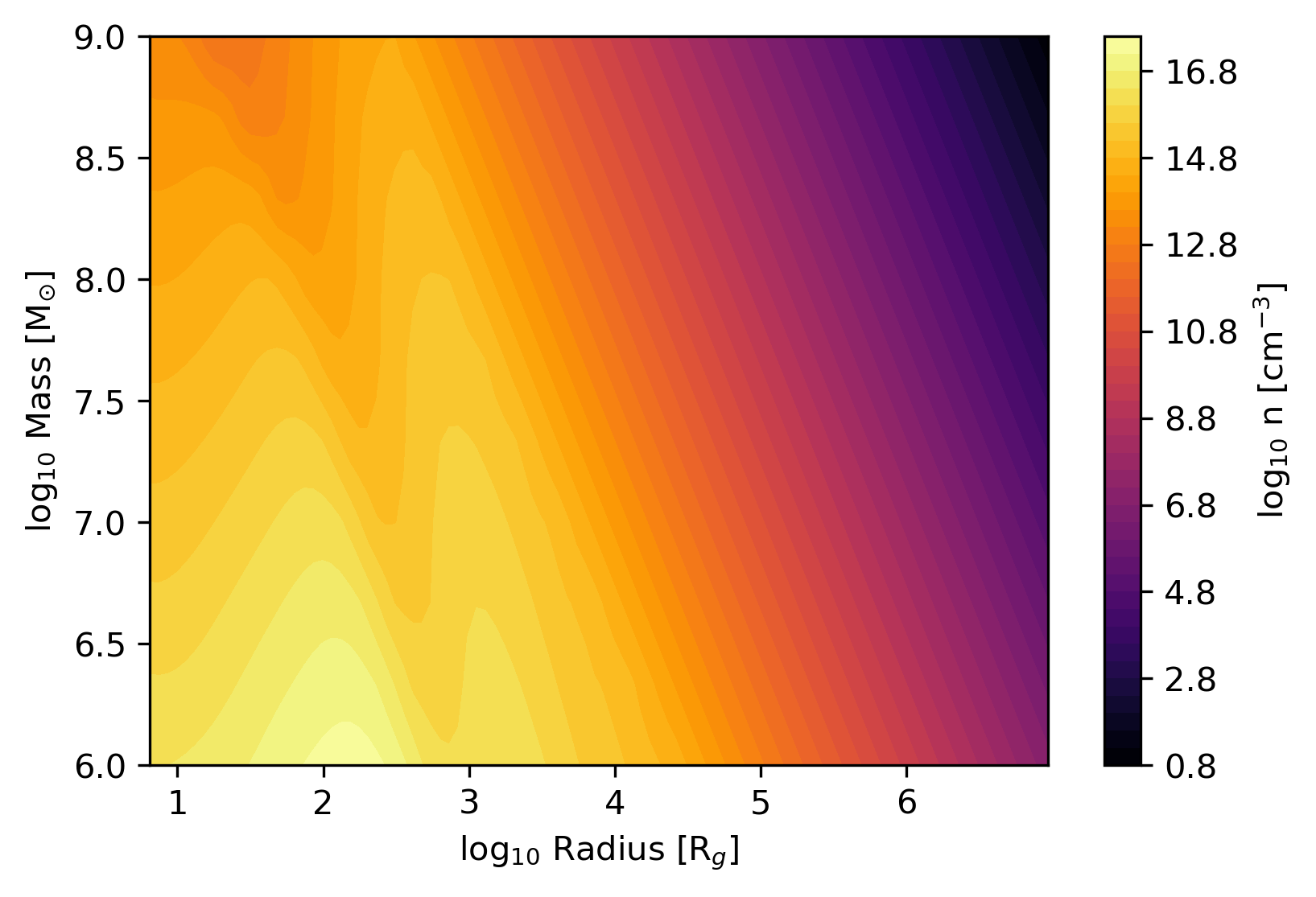}
     \includegraphics[width=0.45\textwidth]{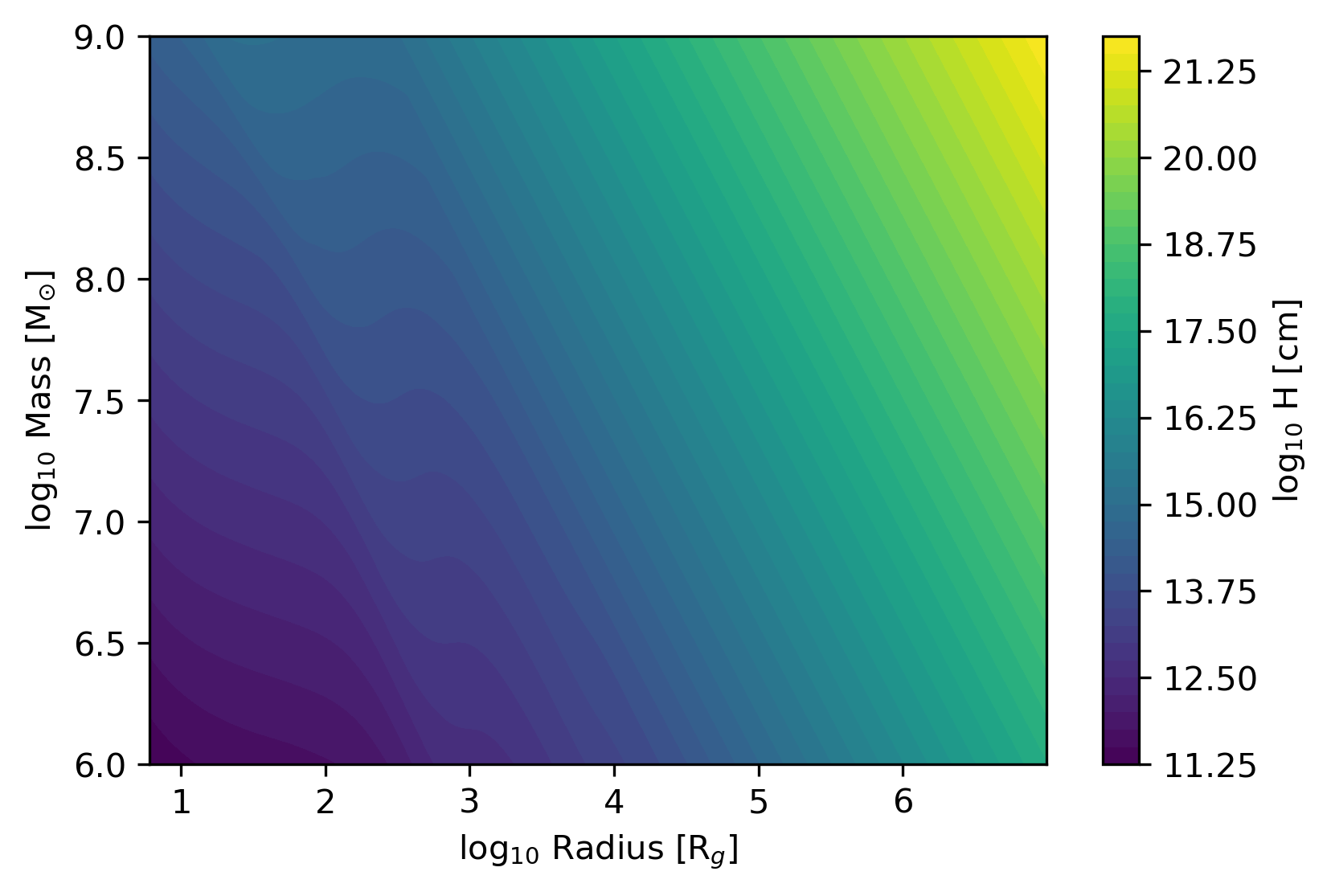}\\    
     \includegraphics[width=0.45\textwidth]{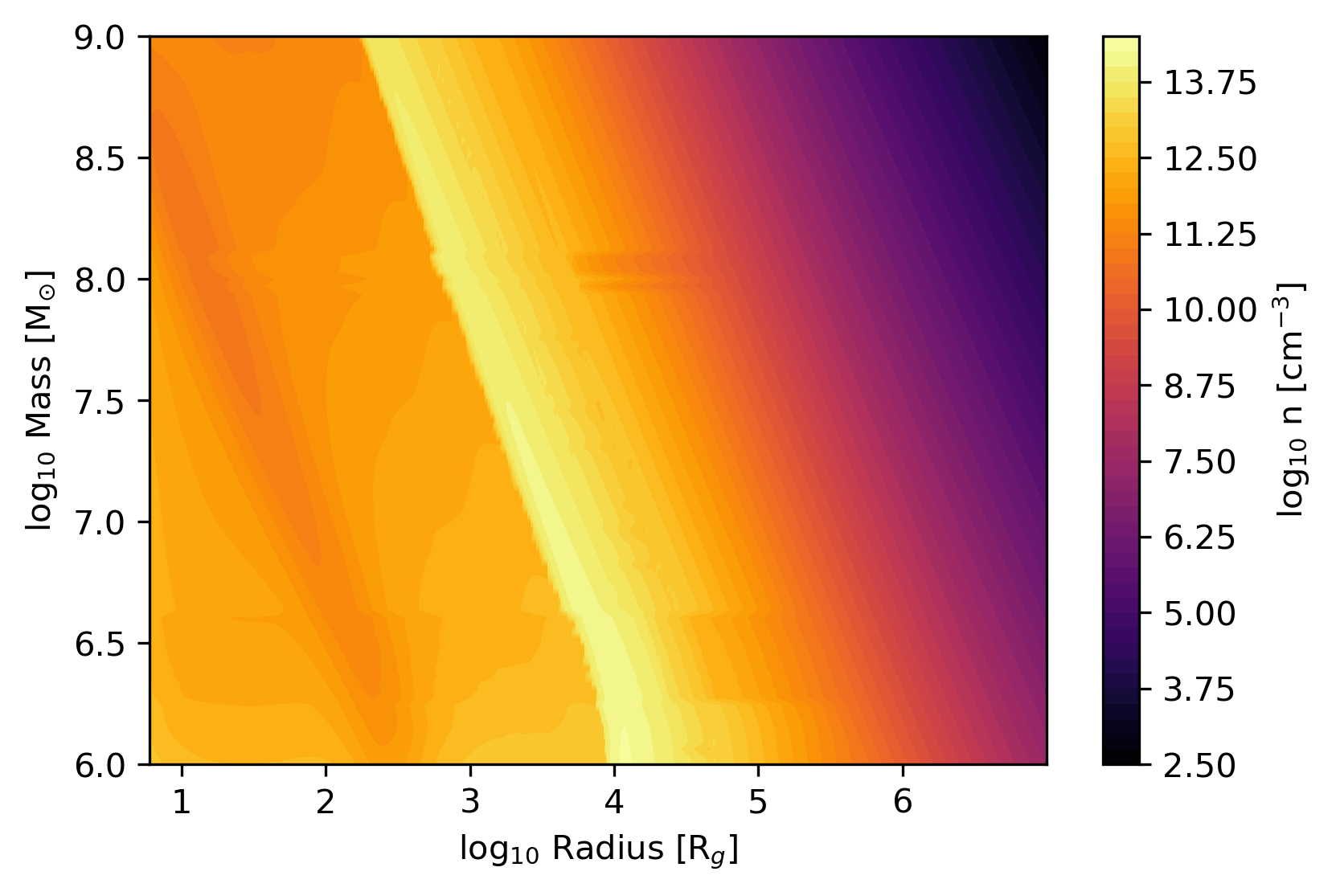}
     \includegraphics[width=0.45\textwidth]{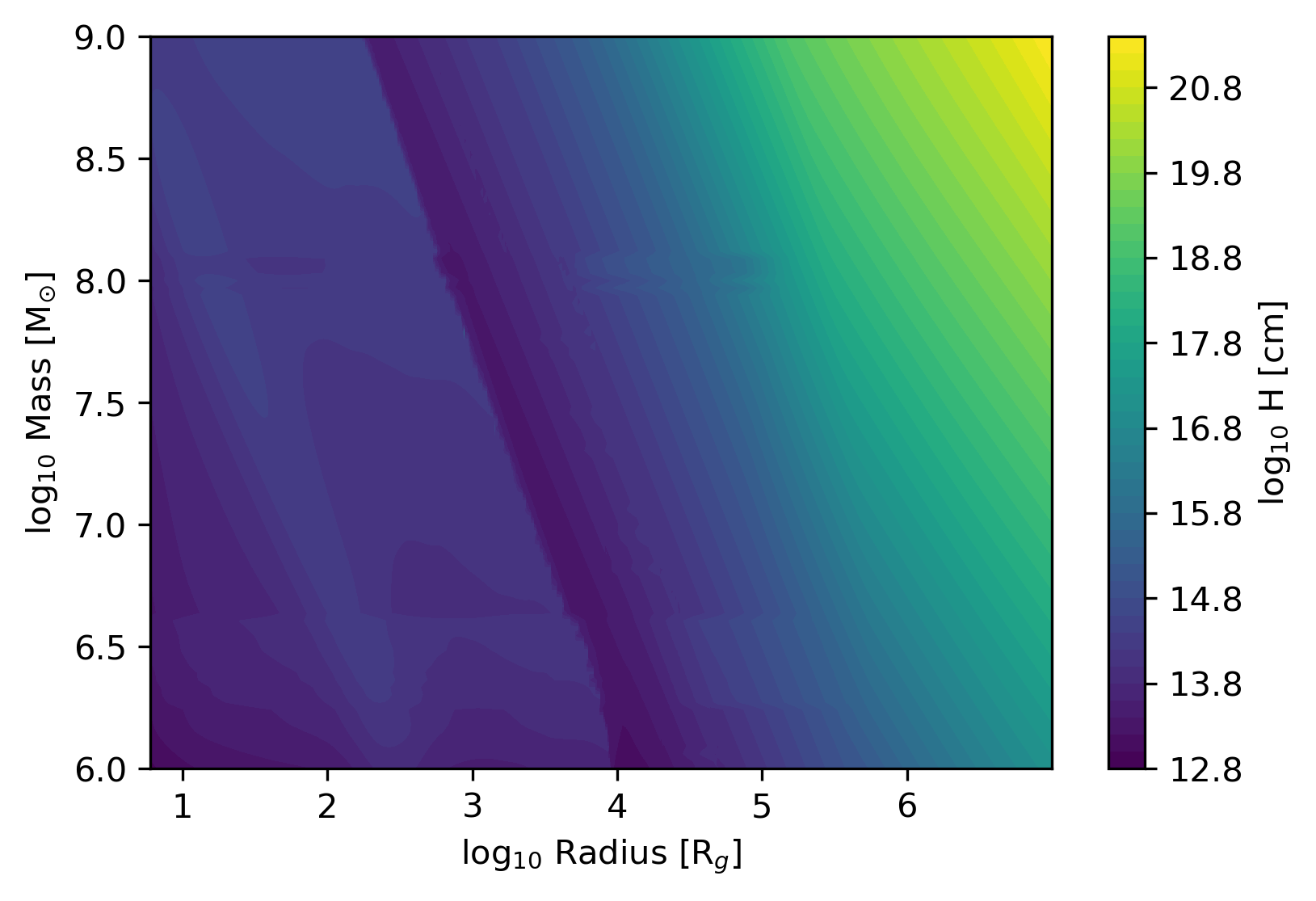}
    \caption{\small {
    Properties of the SG (top panels) and TQM (bottom panels) disk models adopted from pAGN code. Left figures show the {disk mid-plane} number density, $n$, contour map as a function of radius and SMBH masses in the range of $10^6$ - 10$^9$ M$_{\odot}$. Similarly, right figures show the scale height, $H$.}}
    \label{fig:disk}
\end{figure*}

\begin{figure*}
    \centering
    \includegraphics[width=0.45\textwidth]{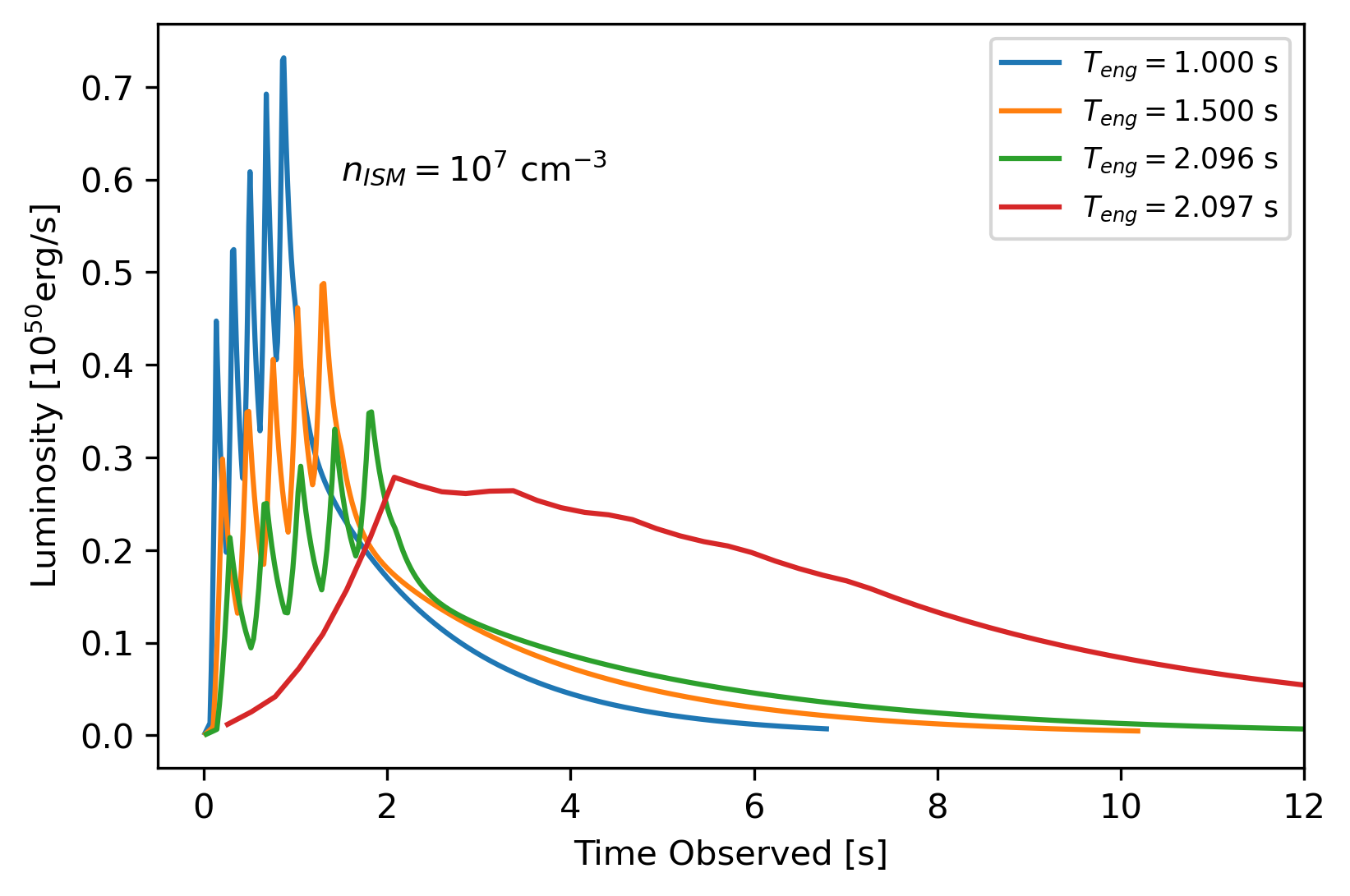}
     \includegraphics[width=0.45\textwidth]{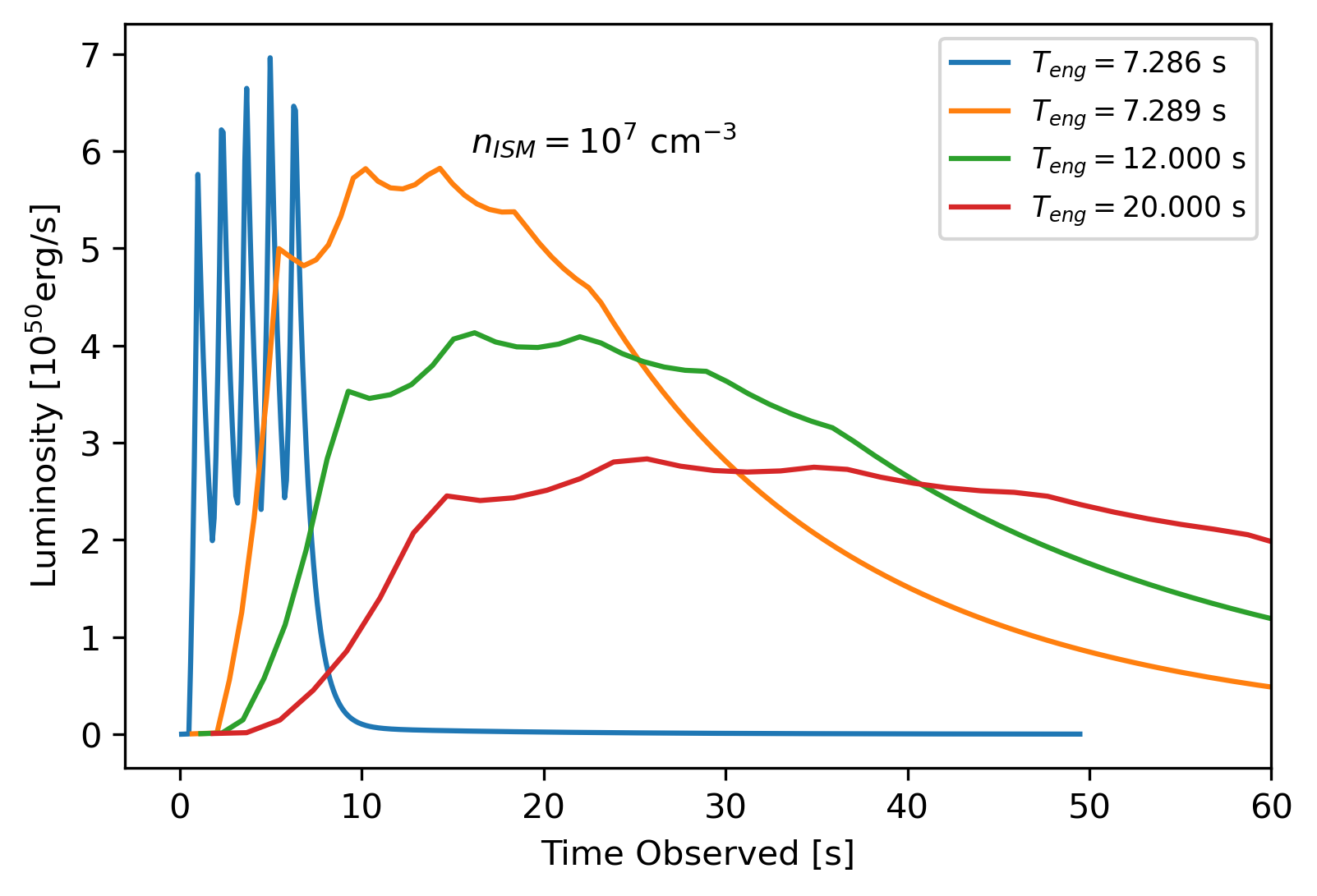}\\
       \includegraphics[width=0.45\textwidth]{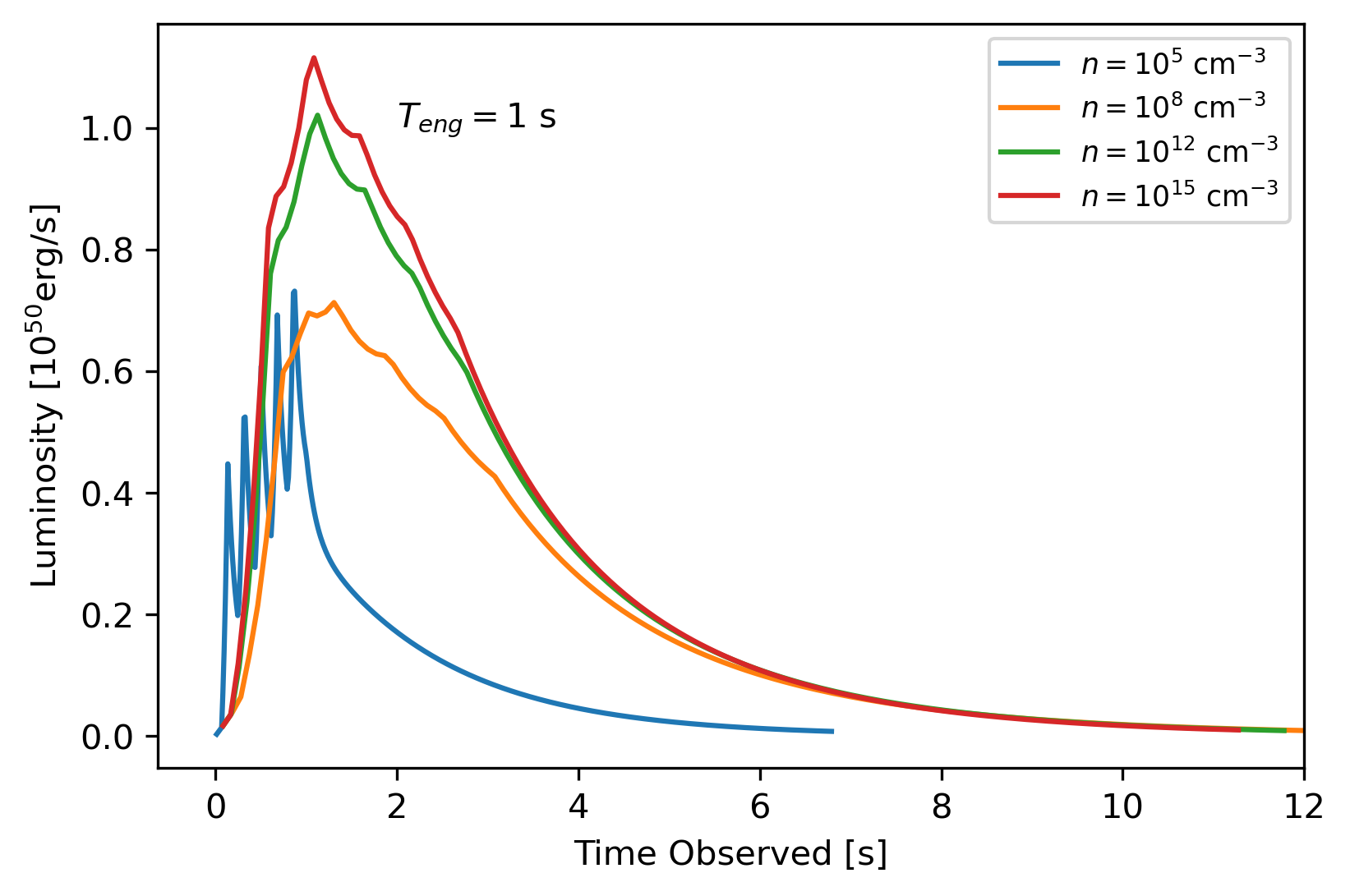}  
       \includegraphics[width=0.45\textwidth]{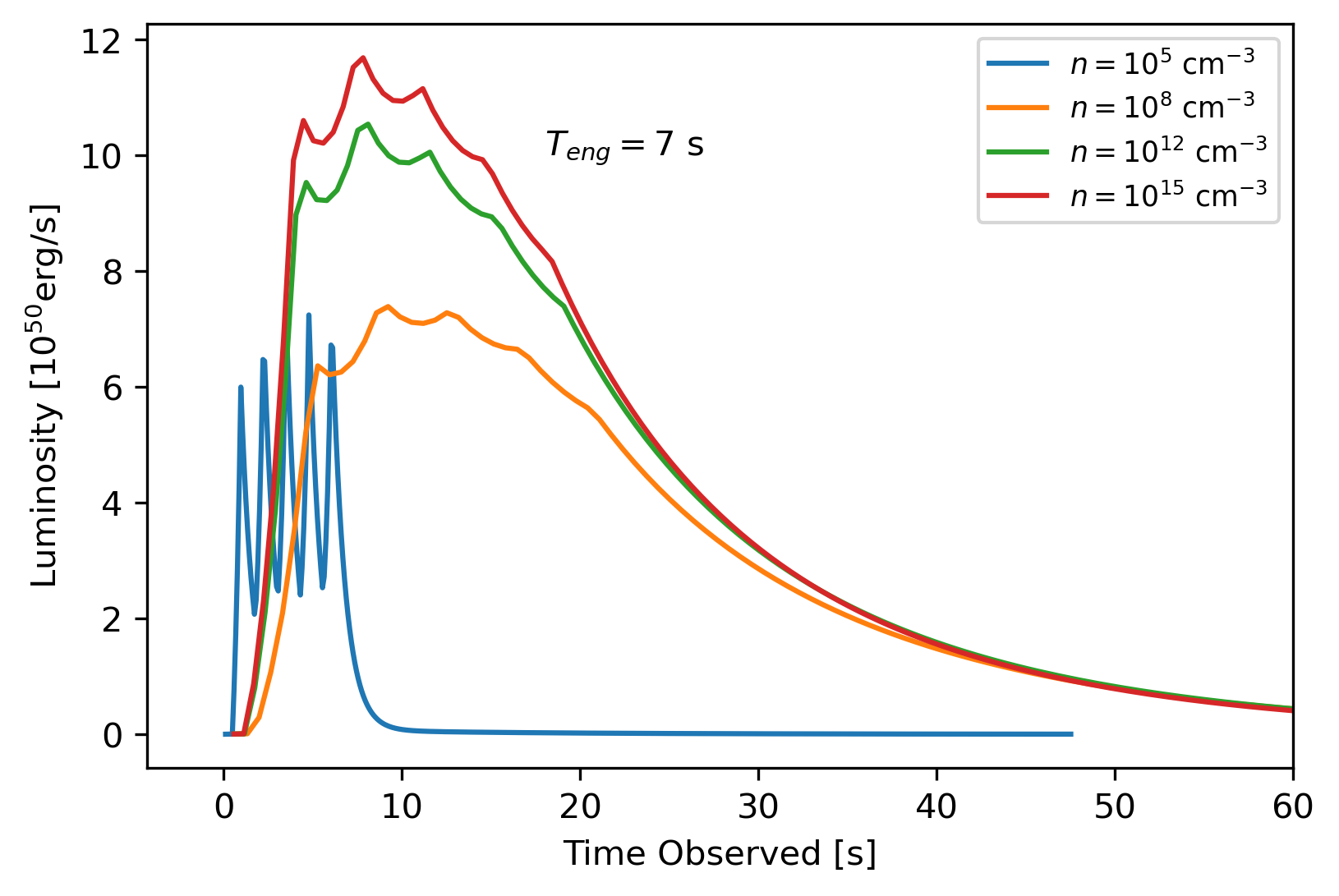}       
    \caption{\small Simulated light curves for a short-engine (left panels) and a long-engine (right panels) HD-GRB.
    In both cases, the engine has 5 distinct radiation pulses.
    The top panels show the variation with engine duration at fixed ambient density, while the bottom panels show
    the variation with density for a duration of the engine. In all the cases 
the light curves are characterized by a single FRED pulse, resulting from the superposition of individual 
pulses. This causes variability overlaid on the broad pulse envelope. 
 The effective duration of the burst increases in all the cases, remarkably making also the short GRB to appear like a long one.
 The GRB microphysical parameters used in the light curve simulations are:  $E_{\rm iso, c}=10^{53}$~erg for the LGRB and 
$E_{\rm iso, c}=10^{51}$~erg for the SGRB, with $\Gamma_\infty=100$  in both cases.}
    \label{fig:prompt}
\end{figure*}

\section{Methods}
\label{sec:methods}

This section describes the various ingredients of the modeling, from the cosmological
population of AGN disks, to the specific disk structure, to the GRB emission properties
in  AGN disk environments.

\subsection{The cosmological population of AGN disks}

Our modeling assumes a direct link between the mass of a SMBH and the accretion disk
feeding it, as typical of accretion disk models (see \S2.1.2). Hence the first element of the
modeling is the cosmological distribution of SMBHs, which will be discussed in \S2.1.1.

\subsubsection{The AGN SMBH cosmological distribution}
The redshift and AGN SMBH mass (connected to the disk mass) are assumed to follow the probability distributions
derived by \citet{Merloni2008} (see also  \citealt{Kelly2012}). 
In this study, we adopt the SMBH mass function profile from the work of \cite{Merloni2008}. This profile assumes a continuous, broad distribution of accretion rates across SMBHs, where every SMBH is active at some level. The most actively accreting SMBHs manifest observationally as AGNs. This approach avoids distinguishing between ``quiet" and ``active" SMBHs, instead positing that the activity level is a function of the accretion rate distribution.
The SMBH mass function density, $\Phi_{M}(z,M)$, is constructed using the average values within the uncertainty bands provided in their work (cfr. Fig.~9 of \citealt{Merloni2008}). 
Using the fits to their models for a range of redshifts between 0 and 5, we then perform a bilinear interpolation to extract the SMBH mass function profile for any value of redshift and SMBH mass as needed by our models. Examples of the mass dependence of the function $\Phi_{M}(z,M)$  at various redshifts are shown in 
Figure~\ref{fig:BHMF}.

\subsubsection{Disk model}
The Shakura-Sunyaev model (\citealt{Shakura1973}, SS in the following) has been remarkably successful in describing properties of accretion disks.  However, 
it is known to have some shortcomings, such as  
being  susceptible to gravitational instability in the outer regions. This has led to the development of increasingly more sophisticated models appropriate to the AGN conditions, such as the one by \citet{Sirko2003} (SG from now on) and the one by \citet{Thompson2005} (TQM from now on),  
which enforce gravitational stability in the outer disk, and more recently by \citet{Gilbaum2022} and \citet{Epstein-Martin2024}, which include heating feedback
in a more self-consistent way.

In this work we  use the SG and the TQM disk models, as they are widely used in the AGN literature, and also easy to obtain profiles for thanks to a 
recent numerical development, pAGN, by \citet{Gangardt2024}, which solves the SG and TQM disk equations for a wide range of initial conditions and parameters.
In addition, pAGN utilizes updated opacity tables which allow for more realistic temperature and density profiles across the disk. However, while adopting these disk models within the pAGN implementation, we need to keep in mind that even the more sophisticated disk models are still an approximation to reality. 

 The SG model has five input parameters: SMBH mass, accretion rate, radiative efficiency, viscosity coefficient, and pressure flag. The accretion rate is a dependent input derived from the Eddington ratio $l_{\rm E}=\epsilon_s\dot{M}c^2/L_{\rm Edd}$,
where $L_{\rm Edd}$ is the Eddington luminosity and $\epsilon_s$ is the radiative efficiency. The viscosity coefficient, $\alpha$, is a free parameter and independent of other physical properties of the disk as in the standard SS model.
Observations constrain this parameter typically to the range $\alpha \simeq 0.01-0.1$. Here, to be consistent with the value assumed in the standard SG model, we adopt $\alpha = 0.01$, $\epsilon_s = 0.1$, and $l_{\rm E}=0.5$. Thus, the accretion rate, $\dot{M}$, scales directly with the SMBH mass. The pressure flag, $b$, is a power index for viscosity-gas pressure relation which can only be 0 or 1. For our study, we follow the standard SS model and assume $\alpha$-disk by setting $b=0$.

The TQM model has six input parameters: SMBH mass, stellar dispersion, star formation efficiency, angular momentum efficiency, supernovae radiative fraction, and accretion rate. The stellar dispersion,  $\sigma$, is calculated using the $M-\sigma$ relation. We adopt the standard TQM model values for star formation efficiency, $\epsilon_T = 0.001$, angular momentum efficiency, $m = 0.2$, and supernovae radiative fraction, $\xi = 1$. Similar to SG, the accretion rate is not independently specified. However, unlike SG, the accretion rate in TQM is computed dynamically based on the disk's physical properties and processes. The calculation begins with an initial accretion rate at the outer disk boundary, which evolves as the model integrates inward, taking into account radial mass transport and the effects of angular momentum transport. The accretion rate at a given radius is determined by:
$\dot{M} = 4 \pi r \Omega_T m_T \rho H^2$
where $\rho$ is the gas density, $H$ is the disk scale height, $\Omega_T$ is the rotational velocity, and $m_T$ is the angular momentum efficiency \citep{Thompson2005}. The Toomre parameter, $Q$, is used to evaluate disk stability; regions where $Q \sim 1$ undergo star formation, reducing $\dot{M}$ as mass is converted into stars. The star formation rate ($\dot{\Sigma}_*$) per unit area is given by:
$\dot{\Sigma}_* = \Sigma_g \Omega_T \epsilon_T$
where $\Sigma_g$ is the surface mass density. Through this iterative process, the TQM model dynamically adjusts the accretion rate to account for star formation and feedback effects. 

{Figure~\ref{fig:disk} 
illustrates  disk properties of the SG and TQM models for a range of SMBH masses within $10^6$ - $10^9$ M$_{\odot}$. The left plot shows the disk mid-plane number density, $n=\rho/m_p$ (where $m_p$ is the proton mass), as a function of both disk radius (in units of $R_{\rm g}$) and SMBH mass (in units of M$_{\odot}$). The gravitational radius is defined as $R_{\rm g} = GM_{\rm SMBH}/c^2$, where $G$ is the gravity constant and $c$ is the speed of light.
  For both models, the number density is generally higher in the inner regions and at lower masses, while decreasing at large radii and for higher mass central BHs. The sharp peak around $10^3$ - $10^4$ $R_{\rm g}$ in the TQM model is due to dominant gravitational instability as the Toomre parameter, $Q$, approaches 1. This leads to rapid cooling, allowing gas to accumulate. Beyond $10^4$ $R_{\rm g}$, star formation feedback regulates the disk, causing the density to decline. On the other hand, the SG model gradually approaches gravitational instability, resulting in two moderate density peaks. The inner peak corresponds to the transition from the radiation pressure-dominated regime to the gas pressure-dominated regime. The second peak around $10^3$ $R_{\rm g}$ is the region where self-gravity starts becoming significant, leading to enhanced mass accumulation, but the disk remains only marginally gravitationally unstable. The number density spans a wide range for both models, with the SG disk displaying {a larger range than the TQM due to the steeper scaling in the outer regions.} 
  The right panels of  Figure~\ref{fig:disk} show instead the scale height of the disk, $H$, which can approach $H/R\sim 1$ in the outermost regions, especially at high SMBH masses. A narrow, low scale-height feature is seen in the TQM model, mirroring the high density one discussed above.}

With the disk mid-plane density $n_0(R)$ and the scale height $H (R)$ provided by pAGN, the density profile in the  direction perpendicular to the disk plane, $x$,  can be approximated by an isothermal athmosphere,
\begin{equation}
n(x,R) = n_0(R) \exp \left[-\frac{x^2}{2 H^2(R)}\right]\,.
\label{eq:nz}
\end{equation}

\subsection{GRBs in high-density media}

In the following we describe the changes to the prompt (\S2.2.1) and afterglow  (\S2.2.2) emission induced by
the occurrence of these events in very dense media.

\subsubsection{Prompt emission}
While GRBs originate from compact objects, they are however relativistic sources that 
 become quite extended during their various phases \citep{Piran2004}.
For a canonical burst exploding in the interstellar medium, for example, the gamma-ray phase is produced
 by a fireball of the size of one astronomical unit, while the afterglow radiation develops over length 
scales of light years \citep{Meszaros1997}. For this reason, the observed properties of GRBs are highly 
sensitive to the environment in which their central engines are located (e.g. \citealt{Panaitescu2000}). In particular, there is a critical ambient density at which the hierarchy of internal/external shocks breaks, resulting in bursts with
peculiar characteristics \citep{Perna2021,Lazzati2022}. This happens when the increasing external density
 pushes the onset of the external shock inward until it overlaps with the internal shocks radius. Following \citet{Lazzati2022},
  we define as {High-Density GRBs (HD-GRBs) those bursts for which
\begin{equation}
    n_{\rm{ambient}}>6\times10^6 \, E_{52} \, \Gamma_{\infty,2}^{-8} \,\,
    \Delta{t}^{-3} \,\,\,{\rm cm}^{-3}\,,
    \label{eq:critn}
\end{equation}
where $E_{52}$ is the isotropic equivalent energy of the fireball in units of $10^{52}$~erg, $\Gamma_{\infty,2}$ is the asymptotic Lorentz factor of the fireball in units of $100$, and $\Delta{t}$ is the duration of the engine activity in seconds. In these bursts, the external shock \citep{Meszaros1997} develops  before the internal shocks \citep{Rees1994} take place, giving rise to a burst for which the distinction
 between the prompt and afterglow emission fades.
HD-GRBs are characterized by unique features in their prompt emission. The onset time of each prompt pulse is driven by the Lorentz factor of the ejected fireball \citep{Piran2004,Zhang2004}. The pulse width, instead, is set by the Lorentz factor of the fireball after shocking with the ambient material. As a consequence, the light curve of the prompt emission  is a single, broad pulse made by the superposition of individual ones \citep{Lazzati2022}. 


\begin{figure*}
    \centering
    \includegraphics[width=0.45\textwidth]{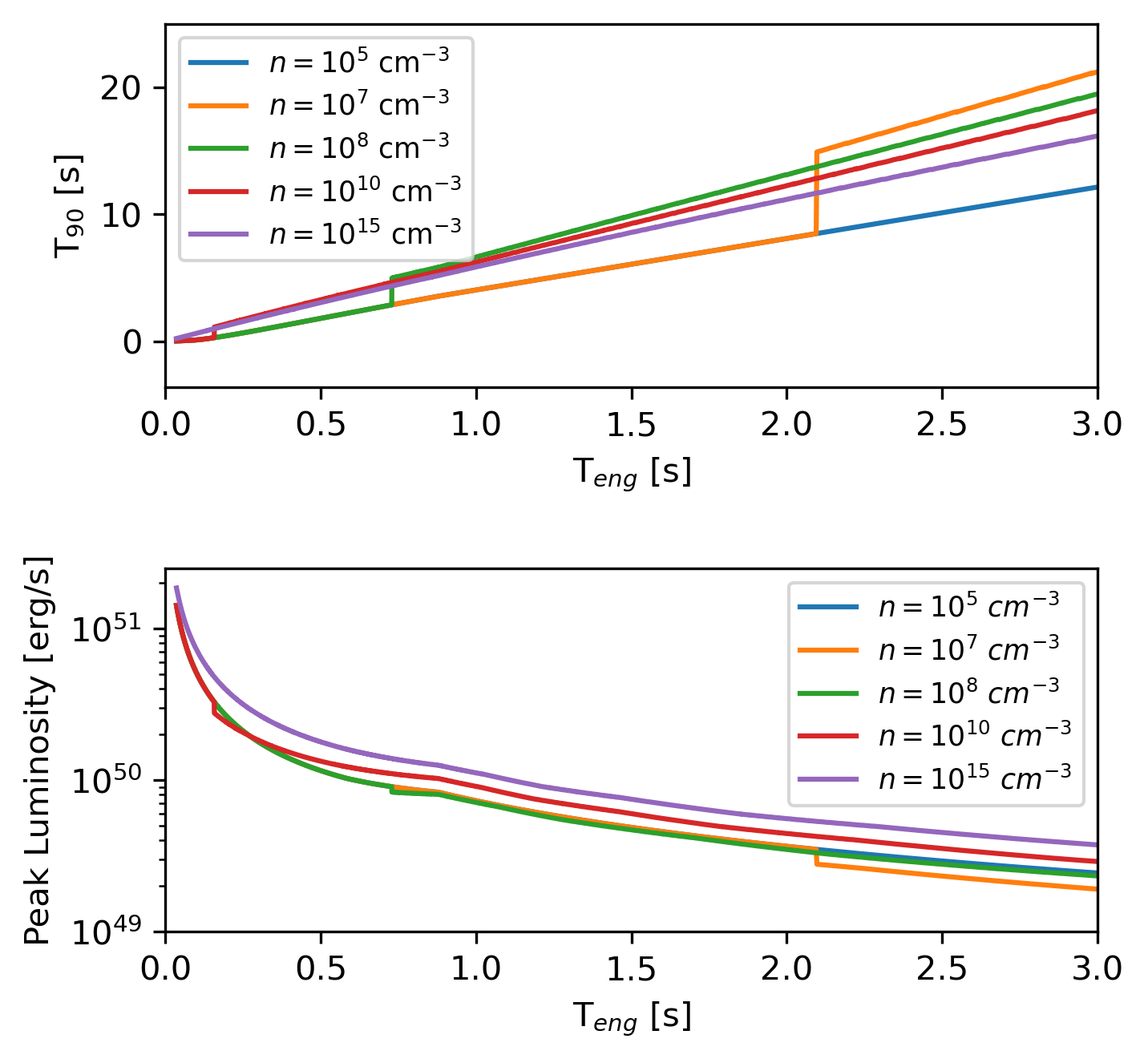}
     \includegraphics[width=0.45\textwidth]{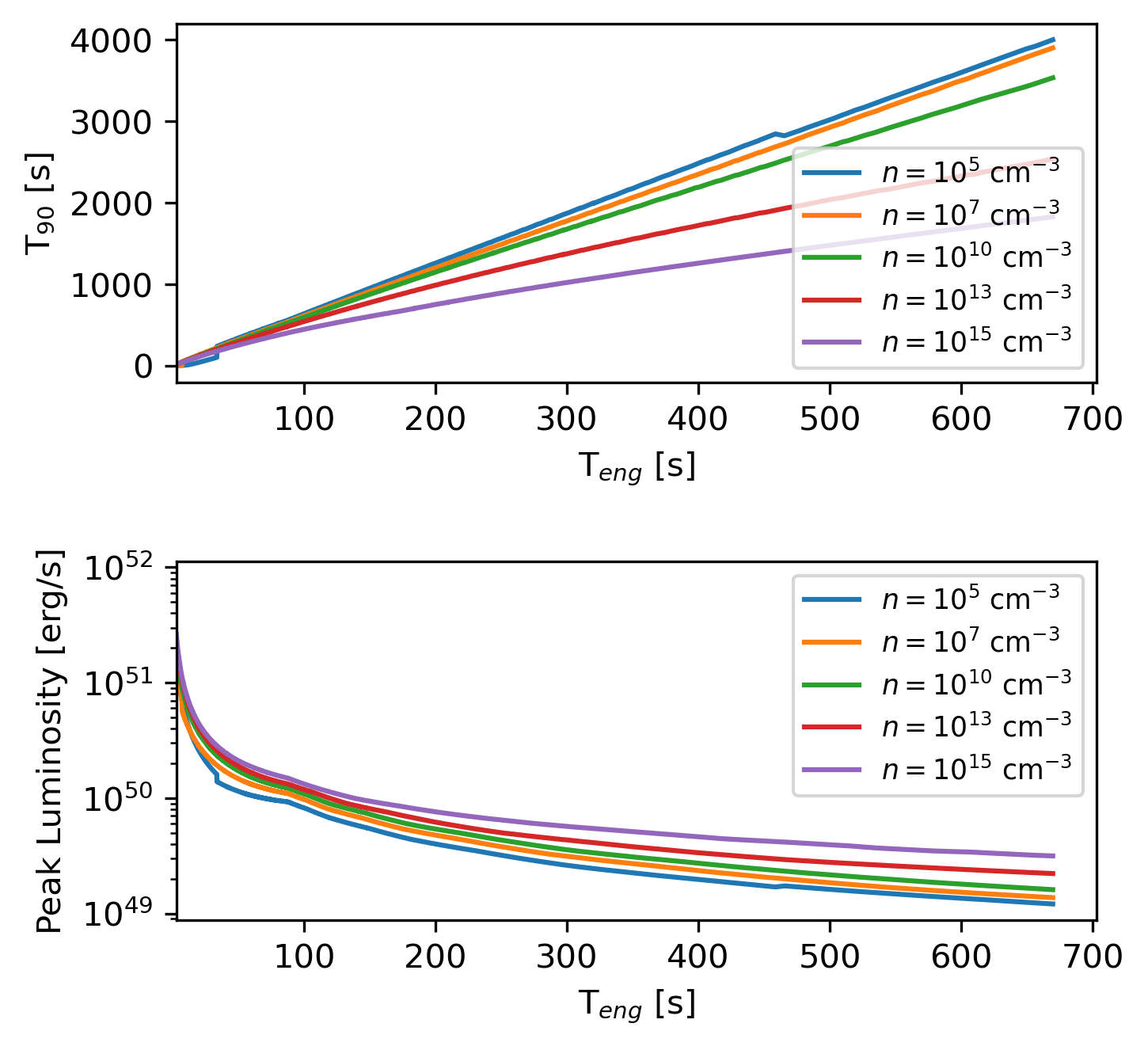}\\      
    \caption{\small {Observed duration $T_{90}$ (top panels) and peak frequency of the light curves (bottom panels)
    as a function of the engine duration (short in the left panels and long in the right ones), for GRBs in high-density media, for various values of the density. 
    For both short and long engines, the time duration of the prompt emission is stretched due to the evolution of the outflow in a high-density medium.
     Jumps are due to the transition from an internal-shock dominated emission, to an external-shock dominated one. }}
    \label{fig:prompt2}
\end{figure*}

\begin{figure*}
    \centering
    \includegraphics[width=0.45\textwidth]{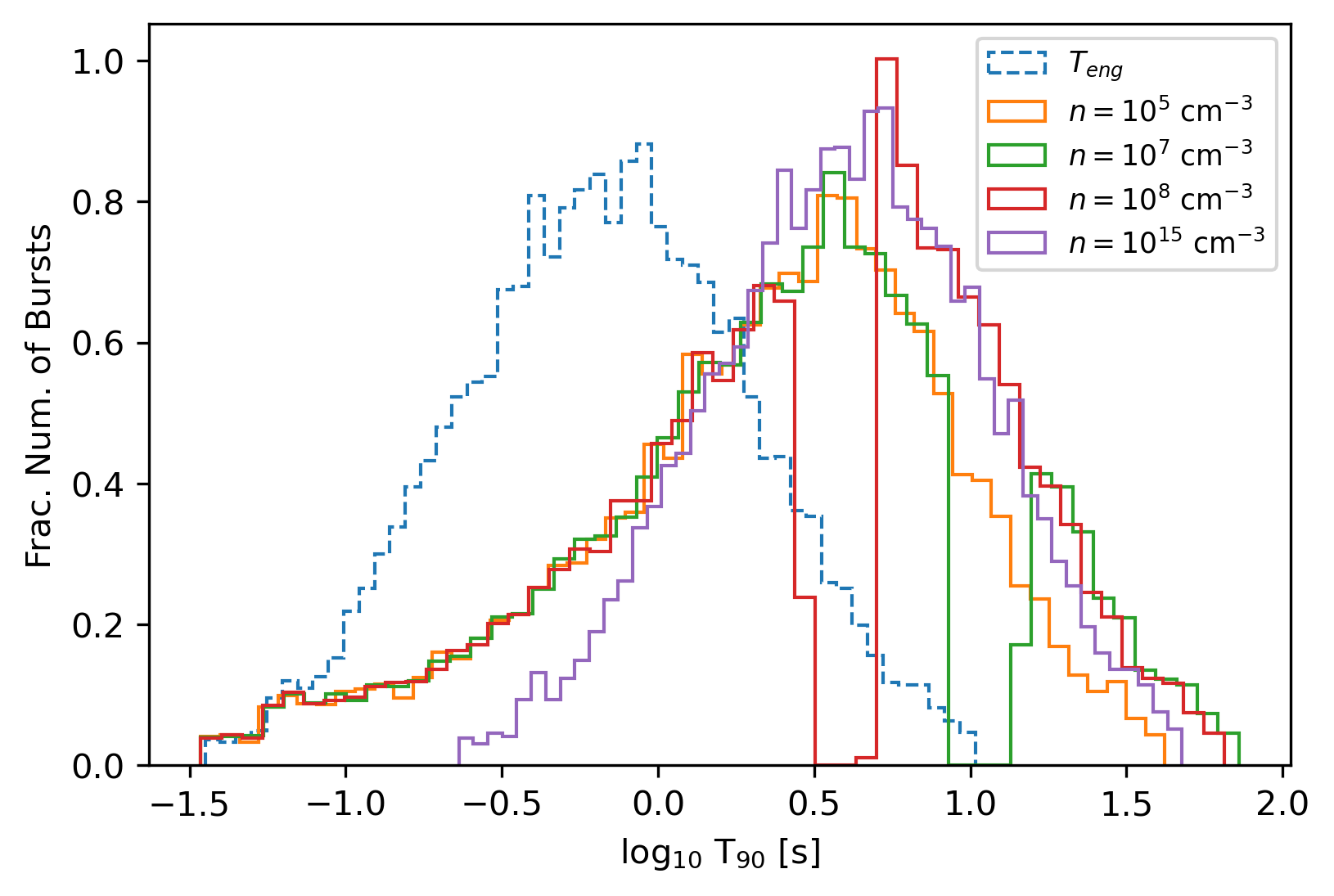}
     \includegraphics[width=0.45\textwidth]{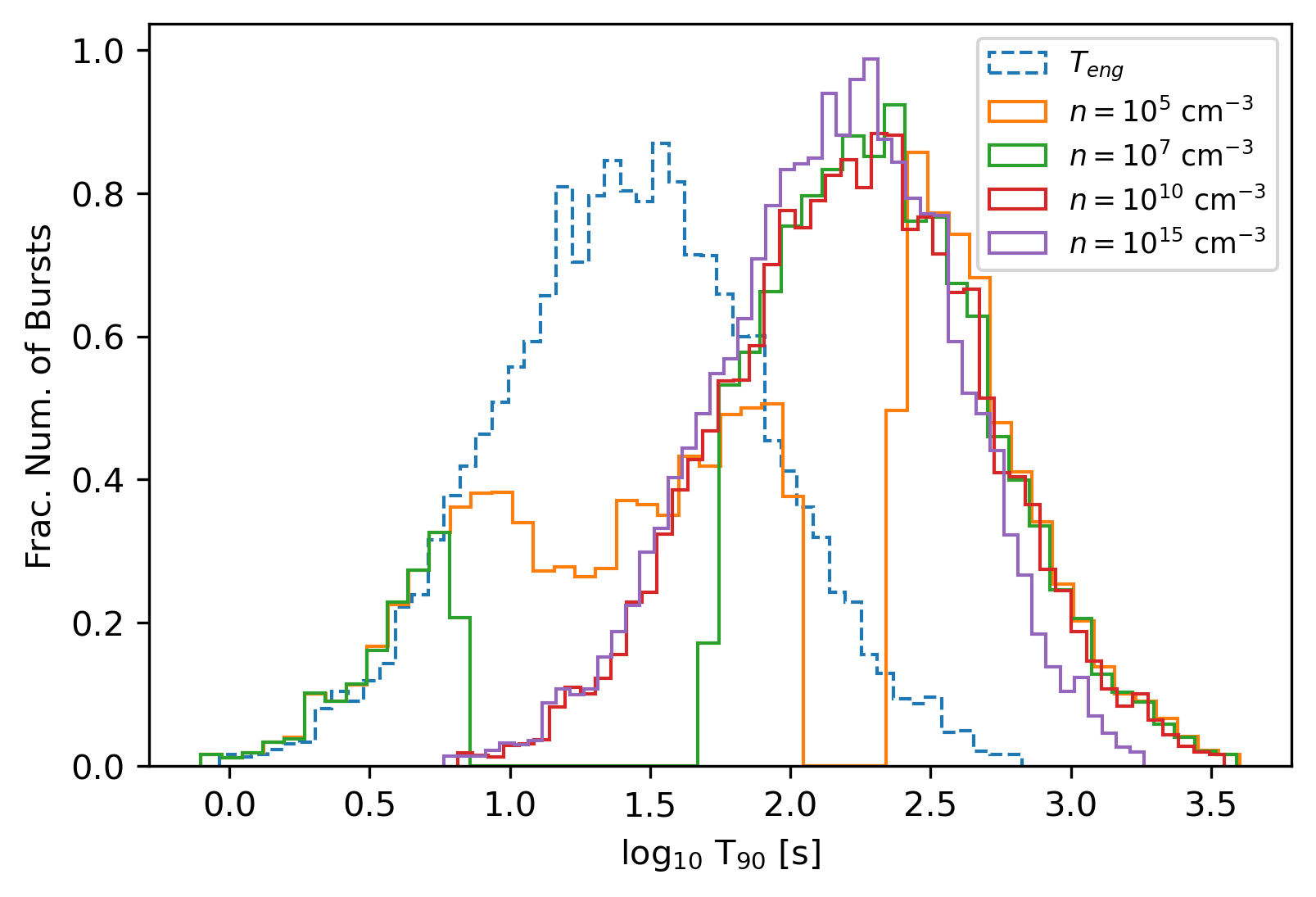}\\      
    \caption{\small {
Probability distributions of event durations of short- (left) and long- (right) duration engines, for GRB sources in high-density media. In both cases, the distribution of engine duration is taken from the observed ones.  Note the shift to longer durations as the density grows, with gaps in the distributions resulting from transition from an internal-shock dominated emission to an external-shock dominates one.} }
    \label{fig:hist}
\end{figure*}

\begin{figure*}
    \centering
    \includegraphics[width=1.02\textwidth]{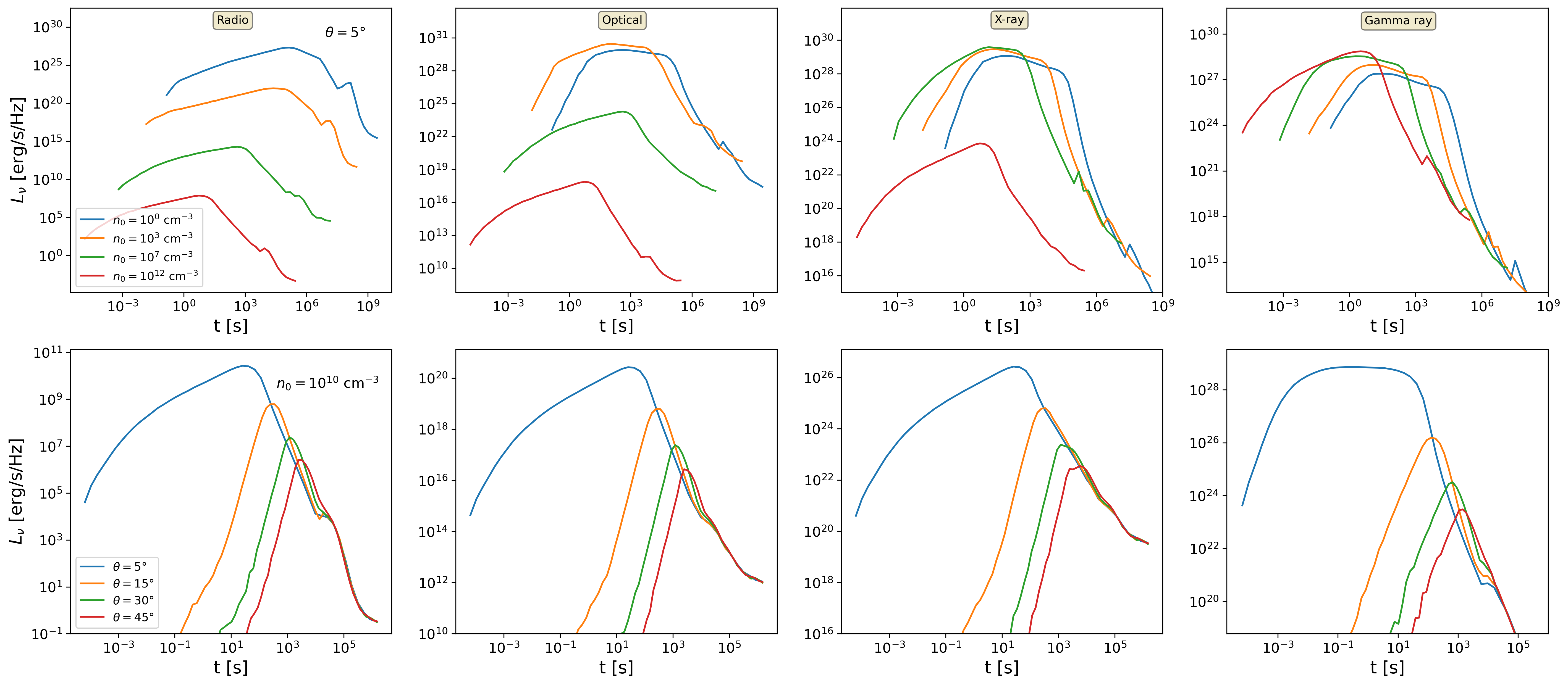}      
    \caption{\small {{\rm Top:} Face-on  afterglow light curves (viewing angle of $5^\circ$) in four representative bands: radio at $4.5\times 10^9$~Hz, optical at $4.5\times 10^{14}$~Hz, X-ray at $4.5\times 10^{17}$~Hz, and gamma ray at $2\times10^{20}$~Hz for a wide range of
densities, from the one typical of the interstellar medium to the high range typical of the discs of AGNs.
{\rm Bottom:} Variation of afterglow light curves as a function of the observer viewing angle, for a fixed density $n=10^{10}$~cm$^{-3}$, as typical of outer regions of AGN disks, from which transients may be more easily detected. Here the GRB isotropic energy is $10^{53}$~erg and the initial Lorentz factor $\Gamma=300$.} }
    \label{fig:aftglow}
\end{figure*}

Figure~\ref{fig:prompt}  shows two sets of simulated light curves, one set for a short GRB engine (left panels) and another for
a long one (right panels), illustrating both the dependence on the engine duration at fixed density (top panels) and on the density for fixed engine duration (bottom panels).  In all cases the code described in \cite{Lazzati2022} was used. The code assumes that the outflow is made of a fixed number of shells (five, in our case) with low Lorentz factor dispersion ($\Delta\Gamma/\Gamma\sim2$). If the external density is above the value specified in Eq~\ref{eq:critn}, the shell dynamics is computed taking into account that the first shell interacts with the static, uniform external medium, while subsequent shells interact with the system made by the decelerated previous shells plus the shocked ambient material. The efficiency is calculated self consistently by conserving energy and momentum and the duration of each pulse is taken to be dominated by the curvature effect. If, instead, the density is below the critical value, the emission is due to collisions between pairs of shells before any interaction with the ambient medium. Collision radii are calculated with basic kinematics and the efficiency is derived from energy and momentum conservation, like in the previous case. Pulse duration is, similarly, dominated by the curvature effect. Each pulse is given a double exponential shape, with the decay phase lasting longer than the rising one.
For the chosen parameters (see caption of the figure), the transition from the standard, 'low-density' GRBs to the HD regime
generally happens between densities of $\sim 10^5-10^8$~cm$^{-3}$, but with a precise value which depends on the engine
duration, since we are assuming a fixed luminosity for all the bursts. 
For both the long and short GRB engines, above the critical density for transition, the light curves display a characteristic fast-rise slow-decay shape similar to the so-called FRED (fast rise, exponential decay) bursts \citep{Fenimore1996}. In addition, some overlaid variability is apparent in the  brightest phases.   Another characteristics found by \citet{Lazzati2022} 
 is a monotonic hard-to-soft spectral evolution \citep{Band1993}, even during phases in which the burst is brightening. Of note, finally, is the fact that 
the short engine produces a burst that, given its duration, would be classified as long. 

Figure~\ref{fig:prompt2} shows the duration of the prompt emission, $T_{\rm 90}$, and the peak frequency evolution for short GRB engines (left panels) and for long GRB engines (right panels) for a range of number densities, $n$, from $10^{5}$ cm$^{-3}$ to $10^{15}$ cm$^{-3}$. 
Similarly to Figure~\ref{fig:prompt}, the transition from the low density regime to the HD regime is observed as a ``jump'' in the trendline for densities between ~$10^{5}$ cm$^{-3}$ and ~$10^{8}$ cm$^{-3}$. This effect is amplified in the short GRB case. At higher number densities, the bursts happen only in the HD regime and the jump is not observed. The lower panels in Figure~\ref{fig:prompt2} show the difference in peak luminosity for the same range of number densities mentioned above. 

A different look at the effect of high medium density on the prompt emission duration is taken in Figure~\ref{fig:hist}.
The probability distribution of the intrinsic engine duration, $T_{\rm eng}$, for both short and long GRBs is assumed to be the same as the $T_{90}$ distribution\footnote{While this equality is not strictly true, it is a necessary approximation here  given that the exact relation varies from burst to burst depending on the spectrum and the observational band and instrumental sensitivity.}, with the latter
taken from observations from the 3rd Fermi/GBM GRB catalogue \citep{Bhat2016}.
Here we are making the reasonable assumptions that the physics of the central engine is not significantly dependent on the density at the
burst location, and that the AGN-GRB sources are subdominant with respect to the global observed GRB population. Hence the properties of this population can be assumed to be a good proxy for the intrinsic properties of the GRB population in AGN disks.
We fit the  \citet{Bhat2016} data using a log-normal distribution to yield the number of bursts, 
$dn/dt_{\rm{eng}}$:
\begin{equation}
   \frac{dn}{d \log\left(t_{\rm{eng}}\right)} = \frac{A}{\sigma (2\pi)^{1/2}}e^{\frac{-(\log \left(t_{\rm{eng}}\right) - \mu)^{2}}{2\sigma^2}}\,. \label{eq:T90}
\end{equation}

The parameters in Eq.~\ref{eq:T90} are shown in Table~\ref{tab:numburst}. The gaps in some of the high-density distributions in Figure~\ref{fig:hist} 
show the transition to the HD regime, and the shift in time duration matches the upper panels in Figure~\ref{fig:prompt2}. As the number density increases, the transition from the standard GRBs to the HD regime happens at shorter burst durations.  
For very high densities,
the evolution of the outflow stays is always in the HD regime for the entire range of engine durations.

\begin{table}[h!]
    \centering
    \begin{tabular}{| c | c | c | c |}
        \hline
        GRB Classification & $A$ & $\mu$ & $\sigma$ \\ \hline
        Short & 36.87 & -0.17 & 0.48 \\ \hline
        Long & 151.71 & 1.43 &  0.48 \\ \hline
    \end{tabular}
    \caption{Parameters of the log-normal distribution fitted to the observed distribution of GRB durations from \citet{Bhat2016}, both for the long and the short GRB populations. 
    }
    \label{tab:numburst}
\end{table}    

\subsubsection{Afterglow emission}

The computation of afterglows in high-density media is performed using
the formalism described in \citet{Wang2022GRB}. The model is based on a relativistic, adiabatic fireball that plows into
the surrounding interstellar medium (ISM). 
A constant fraction
$\epsilon_{\rm e}$ of the internal blastwave energy goes towards
acceleration of electrons into a power-law energy distribution, while a
constant fraction $\epsilon_{\rm B}$ is converted into magnetic
energy.  Under these conditions, the fireball emits synchrotron
radiation, with a broken powerlaw spectrum characterized by three distinct
break frequencies: the injection frequency $\nu_{\rm inj}$ at which new
electrons are injected into the fireball, the cooling frequency
$\nu_{\rm c}$ above which electrons cool efficiently, and the
self-absorption frequency $\nu_{\rm abs}$, below which the synchrotron
radiation is self-absorbed (see \citealt{Sari1998}
for a description of the basic model}).

These frequencies have a different dependence on the density $n$  of the
medium in which the fireball propagates:  $\nu_{\rm inj}$ is independent
of it, $\nu_{\rm c}$ is inversely proportional
to the ambient density, while  $\nu_{\rm abs}$ grows with it
(e.g. \citealt{Panaitescu2000}). This leads to important consequences:
While in typical interstellar media densities the synchrotron self-absorption frequency
$\nu_{\rm abs}$
is smaller than the other two break frequencies, in very high density media it can grow to be
the largest among the three, even reaching the X-ray band.
As a result, the radio and optical/NIR afterglow fluxes are heavily suppressed compared to the  standard afterglow in the ISM. Since the code is designed to provide input for diffusion calculations \citep{Wang2022GRB} the light curves are generated using a Monte Carlo method. Photons are seeded in the external shock according to the local emissivity, isotropic in the comoving frame. Their direction and frequency is then boosted in the laboratory frame. Light curves are then generated by collecting and binning the arrival times of photons in the direction of the selected observer.

Examples of afterglow lightcurves in very dense media can be seen
in Fig.~\ref{fig:aftglow} for {four representative bands: Radio ($10^9$~Hz), Optical ($5\times10^{14}$~Hz), X-rays ($10^{17}$~Hz), and Gamma rays ($10^{20}$~Hz).} The top panels illustrate the dependence on the medium density at fixed viewing angle, for a sightly off-axis GRB, while the bottom panels show the dependence on the viewing angle for a fixed density. In all the cases, the GRB is a representative of a long GRB with isotropic-equivalent energy $E_{\rm iso}=10^{53}$~ergs and initial Lorentz factor $\Gamma=300$. The jet geometry is assumed to be top-hat, with an opening angle of $5^\circ$. Ten million photons were used to generate each set of light curves. Still, the effect of a finite photon number causes the noisy behavior at very low fluxes, easily seen in the upper center and right panels. Notable is the high suppression of the radio flux with density, as a result of the $\nu^2$ low-frequency slope due to self-absorption. {On the other hand, the synchrotron radiation produced by these electrons in gamma rays is above the self-absorption frequency, so the gamma ray flux density is not substantially suppressed \citep{GranotSari2002,Piran2004}.} Note also that the emission peaks at earlier times for higher densities due to the fact the jet break time occurs at earlier times for larger $n$, $t_{\rm peak}\propto n^{-1/3}$ \citep{Sari1999}.

\section{Simulating a cosmological population of long and short GRBs from AGN disks}
\label{sec:simulations}

\subsection{Methods}

We build the cosmological distribution of AGN-born GRBs 
using the model ingredients described in the previous sections, with additional elements detailed below. Since this is the first of such works to date (to the best of our knowledge), we keep our model assumptions to be the simplest.

Our computation evolves through the following steps:\\
{\em (i)} We randomly generate the AGN host redshift through a Monte Carlo method using the SMBH model described in Sec.2.1.1, where the redshift probability distribution is given by
$P(z) = A \int_{M_{\rm min}}^{M_{\rm max}}\Phi(M,z) dM$, and the normalization constant $A$ insures that
$ \int_0^{z_{\rm max}}P(z) dz=1$.  
Note that the underlying assumption of this Monte Carlo realization of the GRB AGN population is that the probability of an AGN disk (of the same mass) to host a GRB does not vary appreciably with the AGN redshift.\\
{\em (ii)} Given the redshift $z_{\rm AGN}$ determined from {\em (i)}, we then randomly draw the mass of the SMBH from the distribution $P(M)\propto M\Phi(M,z_{\rm AGN})$, and assign a disk with properties determined by numerical solutions obtained through pAGN. 
Note that the factor $M$ represents a first order approximation to the assumption that the number of stars and compact objects in a disk is proportional to its mass.\\
{\em (iii)} Given the AGN disk with properties from {\em (ii)}, we randomly draw the radial position $R_{\rm GRB}$ of the GRB within the AGN disk. {Here, for both the case of LGRBS and SGRBs, we adopt a model-independent approach, in which the density of both LGRBs and SGRBs is proportional to the surface mass density at that radius: $P(R)\propto P(\Sigma)$.  By doing so, we assume GRBs are more likely to occur where the surface mass density of the disk is highest, and that the distribution of GRBs is purely dependent on the local surface mass density where disk mass is most concentrated. This approach simplifies the physics to the most basic correlation between disk mass and GRB probability, making it a zeroth-order approximation of the true underlying processes. In future studies, we expect to use this as a baseline for comparison with more sophisticated models that involve more complex physical processes such as the effects of gravitational instabilities on triggering star formation, migration of stars, and feedback mechanisms from stellar physics. In particular, specific modeling of massive stars and of binary NS mergers in AGN disks suggests that migration traps or other dynamical processes may lead to preferential locations for GRB  \citep{Perna2021b,
Fabj2024}. These mechanisms, as well as their dependence on AGN disk properties, are still being investigated in various studies of AGN disks, and therefore we defer to follow up work incorporating them within our formalism.} \\
{\em (iv)} For each population of long and short GRBs, we randomly draw the engine duration $T_{\rm eng}$ from the distribution in Eq.~(\ref{eq:T90}). \\
{\em (v)}  The next step consists in going from the intrinsic distributions of prompt and afterglow radiation computed in {\em (iv)}, to the observable one that emerges from the AGN disk. We will make two assumptions in the following.
The first is that all the sources (i.e. massive stars and neutron stars) are located in the mid-plane of the disk.
While this neglects the possibility that, due to thermal motion,  some of them may be located at various heights within the disk, it is
at the same time the most conservative assumption for our work, in that it considers the maximum absorption at the source.
Secondly, we make the physical assumption that  the direction of the jet is
aligned with the direction of the star spin for the case of massive star collap, and with the orbital angular momentum  of the NS-NS
system in the case of NS-NS mergers.  Next we note that, while in typical galactic environments the direction of the spin of a star and of the orbital angular momentum of an NS-NS binary are randomly oriented, in an AGN disk they are likely to be closely aligned with the orbital angular
 momentum of the accretion disk. This is due to active torques from the disk onto the star \citep{Jermyn2021} in the former case, and to the
orbital  angular momentum alignments of  the NSs orbiting in the disk as they interact to form binaries. {The intrinsic properties of the prompt and afterglow radiation of each source are then computed
using the methods of Sec.~2, using the value of the AGN medium density $\rho(R_{\rm GRB})$.}

Given the location of each burst in the disk, on its way to the observer the radiation will be 
subject to Thomson scattering due to free electrons\footnote{Note that, even if some regions of the AGN disk may be initially neutral,  the early X-ray/UV radiation promptly ionize the medium along the line of sight \citep{Perna2002, Ray2023}, in addition to sublimating dust \citep{Waxman2000, Perna2003}.}, 
resulting in an optical depth 
\begin{equation}
\tau(R_{\rm em},R_{\rm GRB}) = \int_{R_{\rm em}}^\infty  dx\,n(R_{\rm GRB},x) \sigma_{\rm T}
\label{eq:tau}
\end{equation} 
to Thomson scattering, where 
 $n=\rho/ m_p$ is the disk local number density,
and $\sigma_{\rm T}$ the Thomson cross section. 
The variable ${R_{\rm em}}$ in Equation~\ref{eq:tau} represents the radius of the emitting region
of the relativistically propagating jet. As shown in \citet{Perna2021} (see also \citealt{Zhu2021}), depending on the GRB location within the disk,
and on the specific disk model used, the distance over which the jet propagates in the disk from the early, $\gamma$-ray emitting phases,
to the latest radio emitting ones, can be a sizable fraction of the disk scale height. This results in a variety of situations from an entirely
optically thick evolution, to an entirely optical thin, to an intermediate one in which the initial phases have the jet emitting below the disk
photosphere (characterized by $\tau=1$), while the later ones are characterized by the jet emitting after having emerged above
the disk photosphere.

In order to evaluate $R_{\rm em}$ for each source, and for the various phases of the radiation, we need to compute the evolution of the fireball
in the medium. We adopt the modified formulae for thick shells -- which is more general
for the wide range of densities we have here -- developed by \citet{Sari1995,Sari1998} to compute the 
radius $R_{\rm ES}$ of the external shock, defined as the radius at which the fireball dissipates its energy and the afterglow forms
\citep{Meszaros1997}. 

In the traditionally more common case of thin shells, the external shock radius is given by
\begin{equation}
    R^{\rm thin}_{\rm ES} = \left(\frac{3M_{\rm fb}}{4\pi\rho\Gamma_\infty}\right)^{1/3},
\label{eq:Res_thin}    
\end{equation}
where $M_{\rm fb} = {E_{\rm iso}}/{(c^2\Gamma_\infty)}$ is the fireball rest mass.

For a thick shell on the other hand, the thickness $\Delta=cT_{\rm eng}$ plays a role, and the expression
is modified to ensure  that an external shock is formed only after the reverse shock has fully crossed the fireball.
\begin{equation}
    R^{\rm thick}_{\rm ES}= \left(
    \frac{3E_{\rm iso}}{4\pi\rho c^2}\right)^{1/4}\,(cT_{\rm eng})^{1/4}\,.
        \label{eq:Res_thick}
 \end{equation}
To consider here any situation, we take 
\begin{equation}
R_{\rm ES} = {\rm max [R^{\rm thin}_{\rm ES}, R^{\rm thick}_{\rm ES}]}\,.
\label{eq:Res}
\end{equation}      
 
{The prompt $\gamma$-ray emission is produced on very short timescales, during which the fireball typically has not traveled significantly with respect to the disk height. In a high density environment, however, the external shock can form within the timescale of the prompt emission, contributing to the radiation during the early phase. 
Therefore, for both prompt and afterglow emissions, we compute the optical depth in Eq.~\ref{eq:tau} with the emission radius being the radius of the external shock, that is $R_{\rm em} = R_{\rm ES}$.}

{For the ``diffused" scenario, we consider the possible effects that the disk material has on the prompt and afterglow radiation.}
If $\tau \lesssim 1$, the 
GRB radiation will emerge {mostly unaffected}, with the same properties of the intrinsic emission. If, on the other hand, 
$\tau  \gtrsim 1$, the radiation will {diffuse in the disk and emerge} on the
largest between the transient intrinsic duration 
and the
diffusive timescale, which can be approximated as
\begin{equation}
t_{\rm diff} (R_{\rm em},R_{\rm GRB})\approx \frac{[H(R)-R_{\rm em}]\;\tau(R_{\rm em},R_{\rm GRB})}{c}\,.
\label{eq:tdiff}
\end{equation}
Here the quantity $[H(R)-R_{\rm em}]$ represents the
distance that the photons, emitted at a distance $R_{\rm em}$ above the disk plane,
still need to travel within the disk before emerging.
Diffused transients, which emerge on the diffusive timescale, will be observed with a reduced luminosity
$L_{\rm diff}$ compared to the intrinsic luminosity $L_0$. By energy conservation we can write
\begin{equation}
L_{\rm diff} \sim L_{0}\, \frac{t_0}{t_{\rm diff}}
\frac{\Omega}{4\pi}\,,
\label{eq:Ldiff}
\end{equation}
where $t_0$ is the intrinsic duration of the transient and $\Omega$ the solid angle within which the intrinsic emission is initially beamed.
On the other hand, the undiffused 
fraction of the emission which emerges without being scattered is given by 
\begin{equation}
L_{\rm att} = L_{0}\exp(-\tau).
\label{eq:Latt}
\end{equation}
Depending on the value of $\tau$ and the diffusive timescale, the peak luminosity may come from either one of the components in Eq.~(\ref{eq:Ldiff}) or 
Eq.~(\ref{eq:Latt}). Therefore, in our simulations we compute both quantities, and, for each source, we take the peak luminosity to be 
\begin{equation}
L_{\rm red} = {\rm max [L_{\rm diff}, L_{\rm att}]}\,.
\label{eq:Lred}
\end{equation}

The diffusive model for the luminosity described by the previous
equations assumes that there is no mechanical feedback by the stars
and compact objects which ultimately give rise to the long and short
GRBs. However, this is likely not the case for massive objects hosted
in AGN disks.  The very high densities of the AGN disks result in
Bondi accretion rates which are super-Eddington for the majority of
the disk regions.  Under these conditions, winds and outflows are
expected to develop \citep{KingPounds2015}, and possibly carve funnels \citep{Proga2004,Takeuchi2013,Liu2024} which
would then provide a path of lower opacity to the later GRB transient. 
We therefore consider a second model, which we call here ``undiffused",
in which the assumption is that winds from the accreting pre-GRB objects 
have created a path of low optical depth for the GRB radiation to emerge from. {It is noteworthy that there is an overlap between the diffusive and undiffused models, particularly in regions where the optical depth, $\tau$, is less than 1. In such case, the GRB radiation predominantly follows the undiffused model and emerges without significant diffusion.}
We are mindful that the real situation is most likely going to be in between
these two extreme cases, and hence, to be maximally comprehensive,
we provide the results of our simulations in the two scenarios.

{The last remaining variable of our simulated GRB population is the viewing angle $\theta_{\rm obs}$ of the jet axis with respect to the line of sight to the observer. We carry out prompt and afterglow radiation transfer calculations fixing this angle at $\theta_{\rm obs} = 0$ to represent a face-on observation. 
\cite{Perna2021} investigated the role of the viewing geometry on the location of the disk photosphere. They found that the viewing geometry plays a significant role only at large angles, when the line of sight intersects the outer parts of the disc ($\theta_{\rm{obs}}\gtrsim50^o$). At such large angles the disk photosphere increases dramatically, making the transients  undetectable. Since our GRBs are assumed to be beamed within a narrow jet angle, however, bursts at large viewing angles would be intrinsically undetectable even in the undiffused case.}

\subsection{Results}

{In this section we present the results from Monte Carlo simulations of AGN-hosted GRB populations, consisting of $2\times10^5$ and $10^5$ Monte Carlo realizations for the SG and TQM models, respectively.} Redshift, SMBH mass, and location within the disk for each event are generated following the steps described in Sec.~3.1. These properties are assumed to be the same for both long and short GRBs, with the caveats discussed later.

\begin{figure*}
    \centering
    \includegraphics[width=0.45\textwidth]{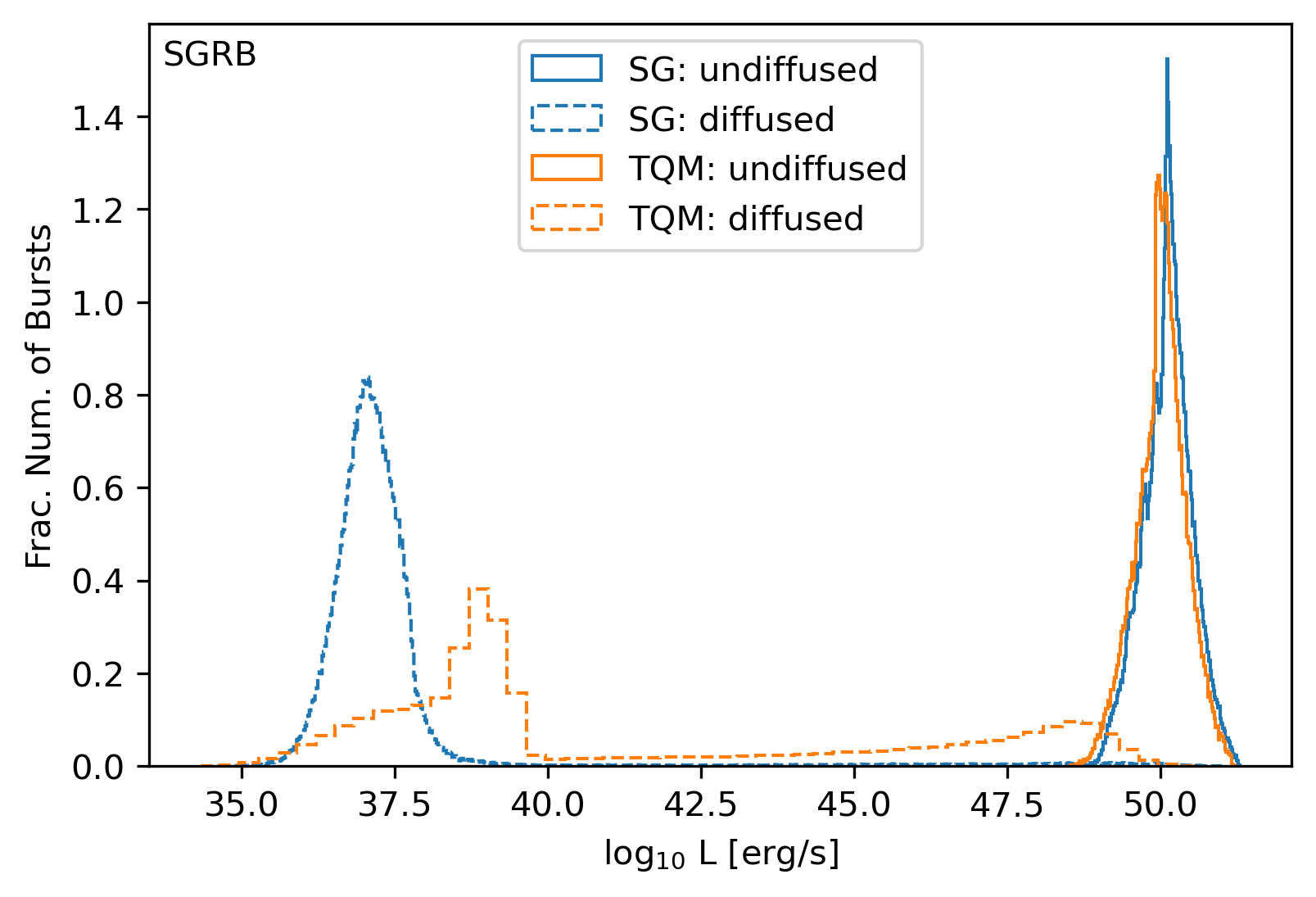}
     \includegraphics[width=0.45\textwidth]{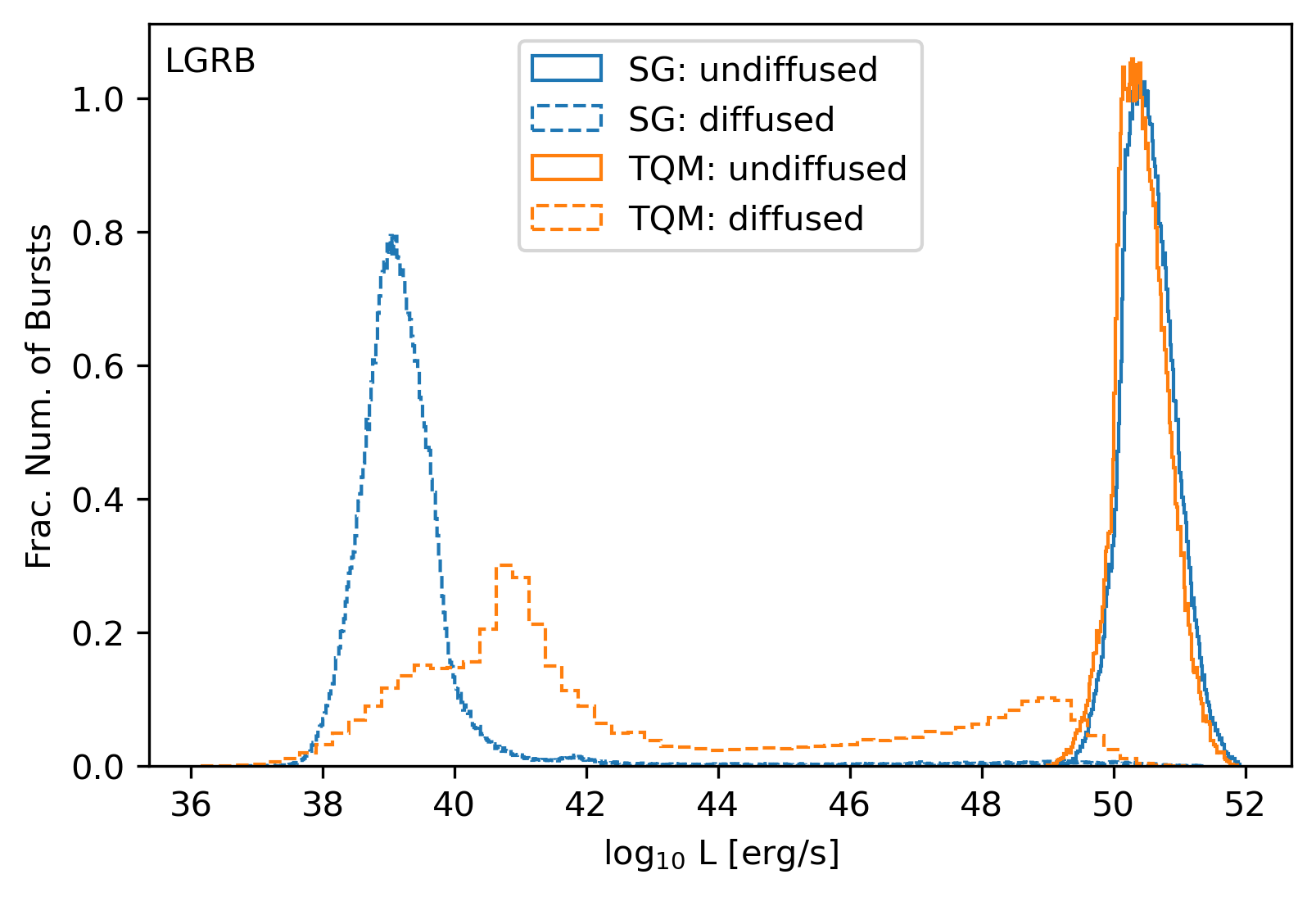}\\      
    \caption{\small{Probability distributions of peak prompt luminosity for SGRBs (left panels)
    and LGBRs (right panels), contrasting the intrinsic luminosity distribution (undiffused model) with the diffused one.
     The former distributions remain relatively narrow, despite some broadening resulting from the stretching in $T_{90}$. Diffusion, on the other hand, results in a considerable dimming and widening of the luminosity distributions.}}
    \label{fig:prmt_lpeak}
\end{figure*}

Fig.~\ref{fig:prmt_lpeak} shows the predicted probability
distribution for the peak luminosity of the GRBs from AGN disks. For
the undiffused case, the probability peaks around $\sim 10^{50}$~erg/s for
SGRBs, and around $\sim 10^{51}$~erg/s for LGRBs. Since, for each
population, we have fixed the explosion energy and the microphysics
parameters regulating the emission, these distributions would
be somewhat narrower for GRBs in the interstellar medium (ISM),
with the spread in peak luminosity being directly correlated to the
burst duration. However, as discussed in the previous sections and
shown in Fig.~\ref{fig:hist}, a wide range of densities results in
a spread of the effective $T_{90}$, and thus of the peak luminosity,
compared to the standard ISM scenario.
The predicted distributions of peak luminosities in the diffusive
scenario are shown in dashed line in Fig.~\ref{fig:prmt_lpeak}.
Diffusion results in heavily suppressed luminosities, by over ten
orders of magnitude for both SG and TQM models. {Diffusion is primarily governed by the optical depth which depends on the disk model. The TQM disk generally has a more uniform optical depth with a higher concentration of values between $10^1$ and $10^2$. In contrast, the SG model features a skewed distribution, with most values between $10^4$ and $10^5$, but also has a larger fraction of optically thin regions, $\tau < 1$. Therefore, the TQM model, characterized by a more uniform distribution of optical depths, results in broader peak luminosity distributions, whereas the SG model predicts a concentration of higher optical depths, resulting in a sharp peak distribution. As shown in Fig.\ref{fig:prmt_lpeak}, TQM has a peak around $L\sim10^{39} \rm erg/s$ for SGRBs and $L\sim10^{41} \rm erg/s$ for LGRBs while SG has a peak at few magnitudes lower, $L\sim10^{37} \rm erg/s$ and $L\sim10^{39} \rm erg/s$, for both SGRBs and LGRBs, respectively.}

\begin{figure*}
    \centering
    \includegraphics[width=1.02\textwidth]{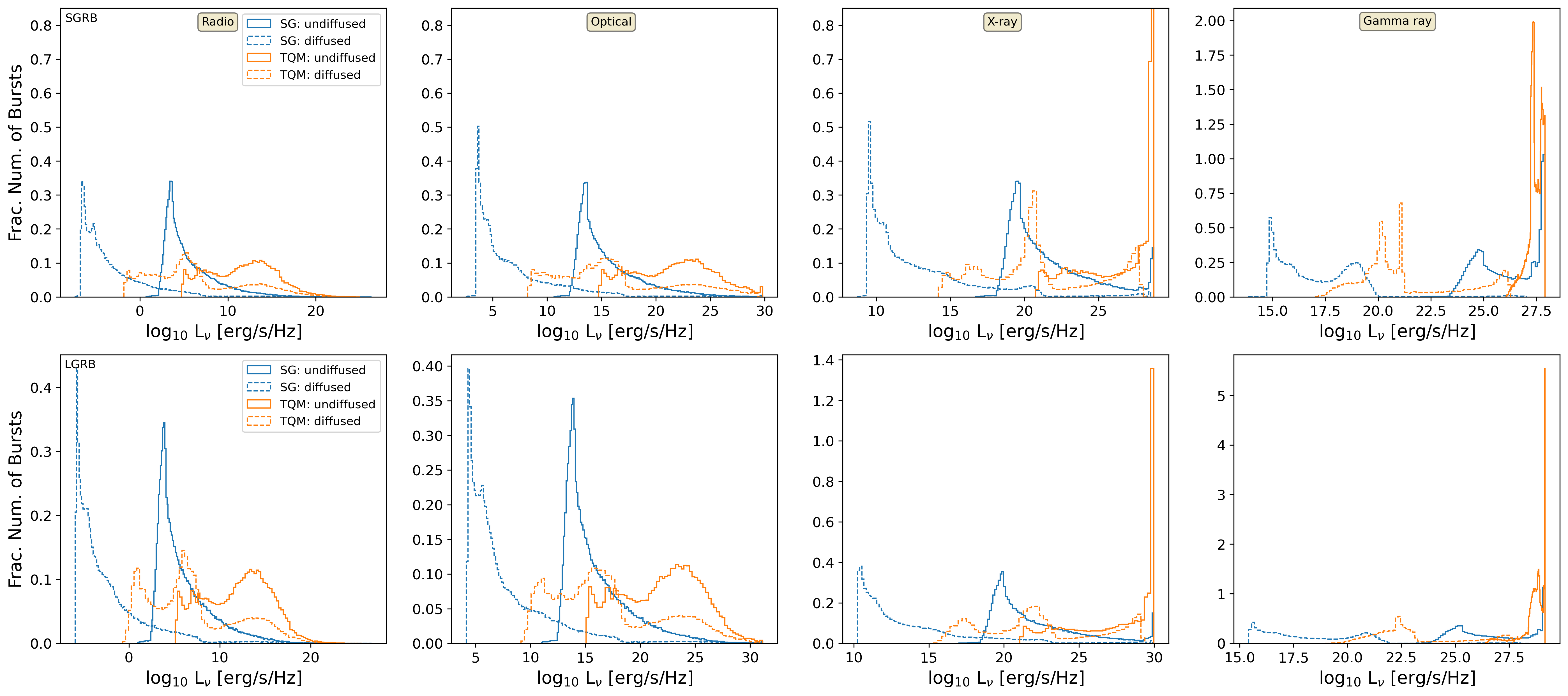}      
    \caption{\small{Afterglow peak luminosity distributions for SGRBs (top panels) and LGRBs (bottom panels) at four representative
      wavelengths: radio, optical, X-ray, and gamma-ray bands.  {For both SGRBs and LGRBs, the peak luminosities in the diffused scenario are several orders of magnitude lower than the luminosities in the undiffused case. The luminosity distributions in the TQM disk, as a result of the typically lower densities than in the SG disk, peak several orders of magnitude higher than they do in SG. The diffused distributions are much broader and dimmer, as expected.
       }}}
    \label{fig:aftglow_lpeak}
\end{figure*}

Similarly to the prompt emission, Fig.~\ref{fig:aftglow_lpeak}
displays the afterglow distributions for both the undiffused and the
diffusive cases. However, unlike the prompt emission, whose intensity
is only marginally affected by the very high densities, the afterglow
can be significantly suppressed especially at the lower wavelengths,
as explicitly shown in Fig.~\ref{fig:aftglow}. This is clearly reflected
in the distributions of Fig.~\ref{fig:aftglow_lpeak} which, even for
the undiffused case, are dominated by low-luminosity events, which
are the ones corresponding to the highest densities.
Note that the highest energies, in $\gamma$-rays, display a bimodal
distribution of luminosity. The peak at higher values corresponds to
relatively lower densities, which have not been much affected by
synchrotron self-absorption, while the highly suppressed second peak
represents the bursts from much denser regions, in which even the
$\gamma$-rays become self absorbed.

{Likewise for the prompt emission, the diffused light curve distributions, depicted in dashed lines in Fig.~\ref{fig:aftglow_lpeak}, are much broader and dimmer than the ones for the undiffused scenario. As the frequency decreases to the X-rays (second panel from the right
in Fig.~\ref{fig:aftglow_lpeak}), the second peak at higher
luminosity decreases and eventually disappears, as the effect of
self absorption becomes increasingly more important for higher
densities. This is consistent with the results from Fig.~\ref{fig:aftglow} where the luminosity is suppressed with increasing density as the frequency decreases. Specifically, the TQM model predicts afterglow peak luminosities that are significantly higher, by several orders of magnitude, compared to those produced by the SG model. This difference arises because the TQM model features a lower number density distribution, with a peak around $10^8 \rm cm^{-3}$, while the SG model predicts higher number densities, with most values around $10^{15} \rm cm^{-3}$. From
the results presented in Fig.~\ref{fig:aftglow_lpeak} it is already
evident that, even in the most optimistic scenario of the undiffused
emission, the best prospects for observing GRBs from AGN disks are in
the prompt emission and in the high-energy component of the afterglow.}

To better quantify the statement above, and especially connect it to
the broadband observability with current facilities at various
wavelengths, we need to compute the distribution of the observed
fluxes, and evaluate the fraction of events that are above the
various detection thresholds.

\begin{figure*}
    \centering
    \includegraphics[width=0.45\textwidth]{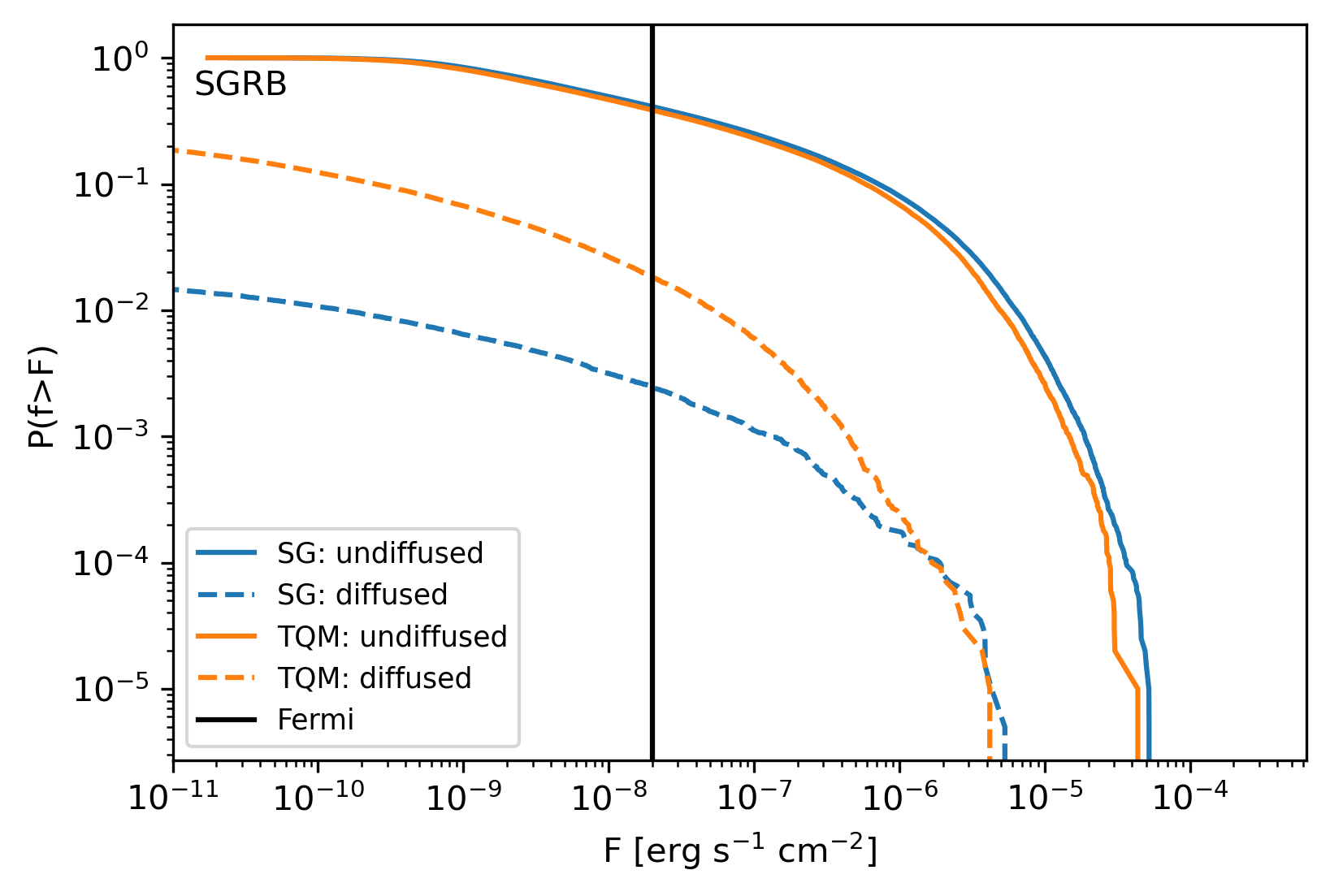}
     \includegraphics[width=0.45\textwidth]{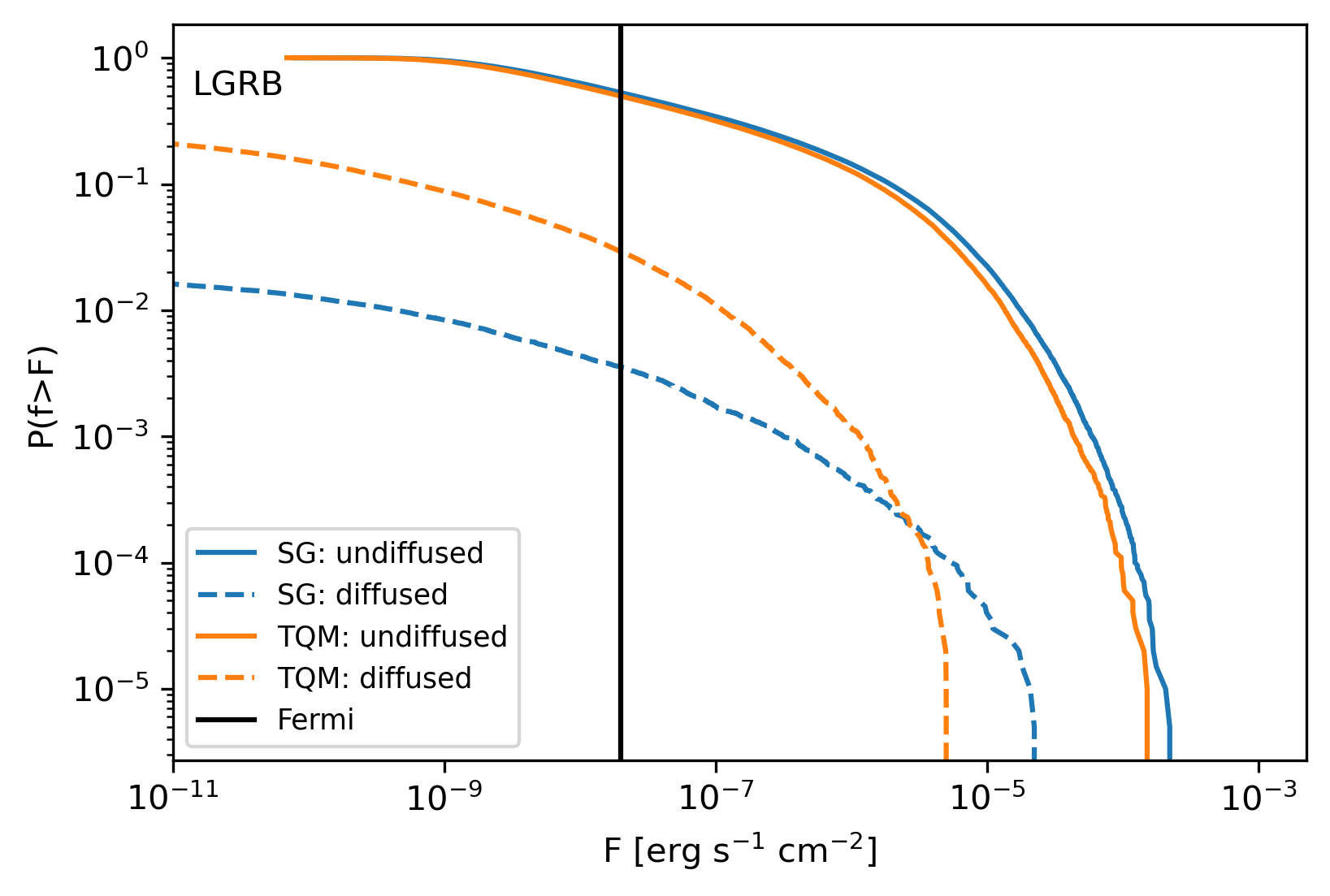}\\      
    \caption{\small{Cumulative distribution function of prompt emission flux for SGRBs (left)
    and LGRBs (right), again contrasting the intrinsic distributions of the undiffused (solid line) scenario with the diffused ones (dashed line). Blue and orange lines represent the curves for the SG and TQM models, respectively.}  
    The vertical black line represents {\em Fermi GBM}'s detection threshold. In both SGRBs and LGRBs, a large fraction of the diffused population is below the Fermi limit and would thus be undetectable.} 
    \label{fig:prmt_ccdf}
\end{figure*}

\begin{figure*}
    \centering
    \includegraphics[width=1.02\textwidth]{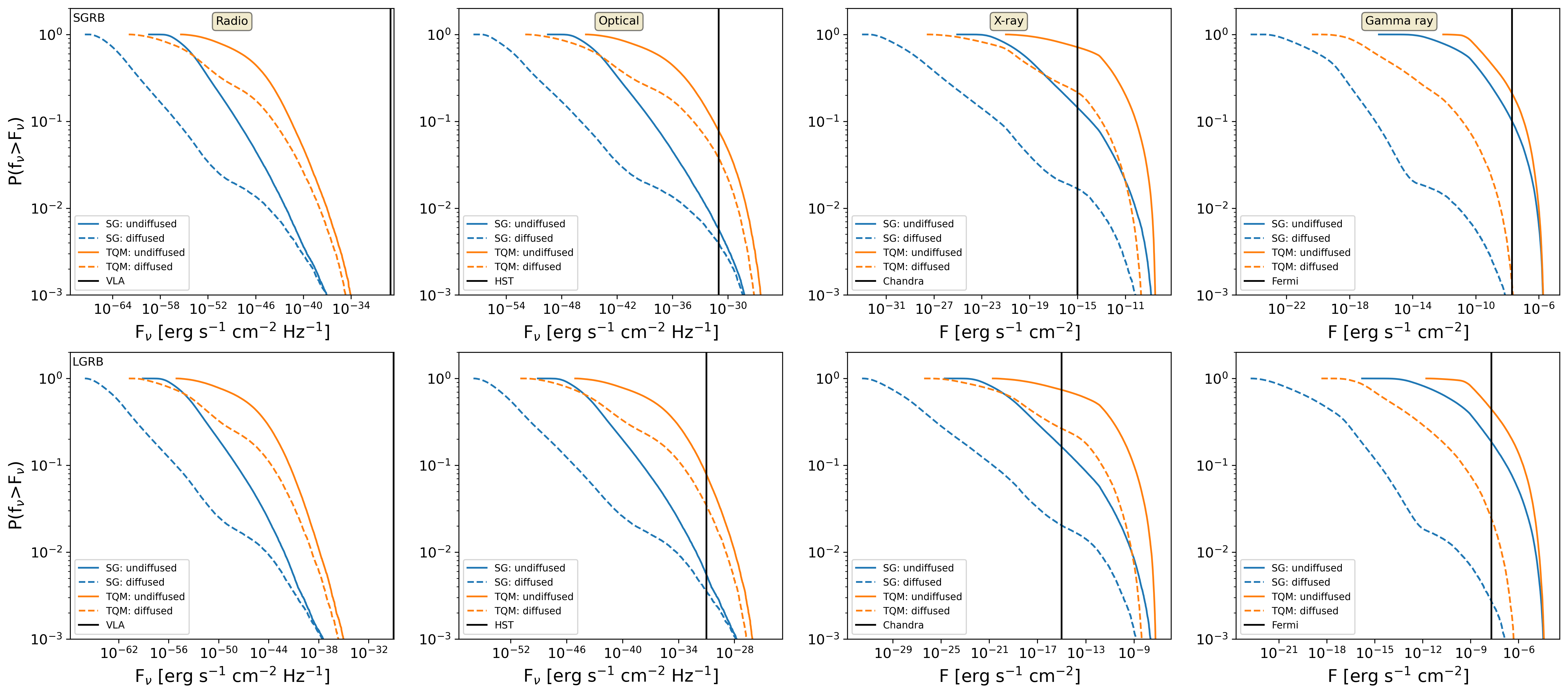}      
    \caption{Cumulative distribution functions for the peak afterglow flux densities of SGRBs (top row) and LGRB (bottom row) across multiple energy bands: radio (left), optical (second from left), X-ray (second from right), and gamma-rays (right). {Similarly to Fig.~\ref{fig:prmt_ccdf}, undiffused (solid line) and diffused (dashed line) distributions are separately shown. Blue and orange lines represent the curves for the SG and TQM models, respectively.} The vertical black lines indicate the sensitivity thresholds for specific observatories and integration times (see text): {\em VLA} for radio, {\em HST} for optical, {\em Chandra} for X-ray, and {\em Fermi GBM} for gamma-ray. {The highest fraction of observable events is present in X-ray for both SG and TQM models. While SG model has a similar fraction in optical and gamma rays, TQM model has a substantially higher fraction in optical than in gamma rays.} The entire intrinsic and diffused populations in radio are below the VLA limit and thus undetectable.}
    \label{fig:aftglow_ccdf}
\end{figure*}

\begin{figure*}
    \centering
    \includegraphics[width=1\textwidth]{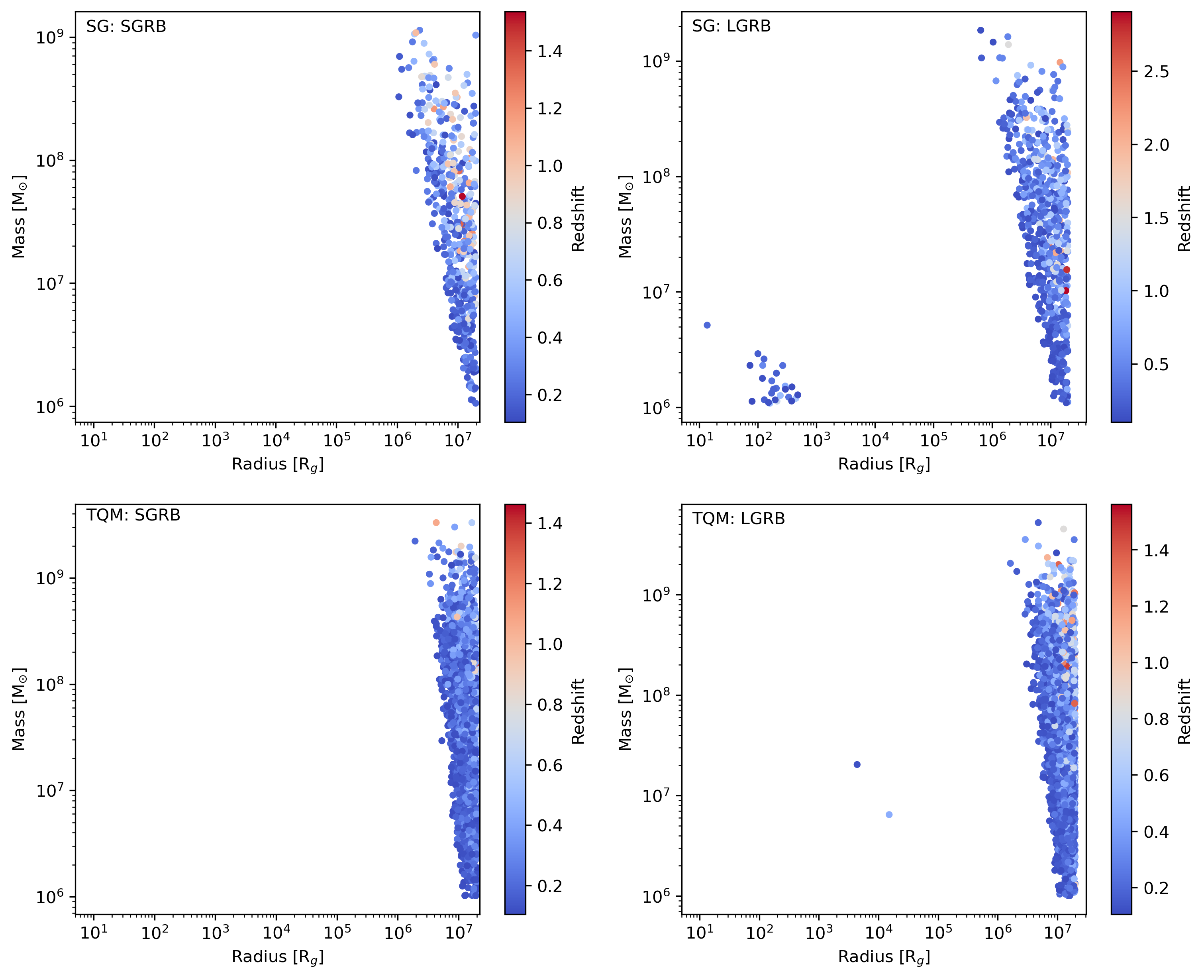}   
    \caption{\small{Scatter plots of mass versus radius for the SGRBs (left) and LGRBs (right) whose prompt emission is above the Fermi/Swift detection sensitivity threshold for SG model (top) and TQM model (bottom). The color bar represents the redshift of the events, where warmer colors (red) indicate higher redshifts and cooler colors (blue) represent lower redshifts. {While most events are detectable at low redshift and within the $R=[10^6,10^7] R_g$ region of the disk, few events in LGRBs in the SG model are detectable at low mass range, $M=[10^{6},10^{7}] M_\odot$ and in the inner region, $R=[10^{1},10^{3}] R_g$.}}}
    \label{fig:prmt_detect}
\end{figure*}

\begin{figure*}
    \centering
    \includegraphics[width=1\textwidth]{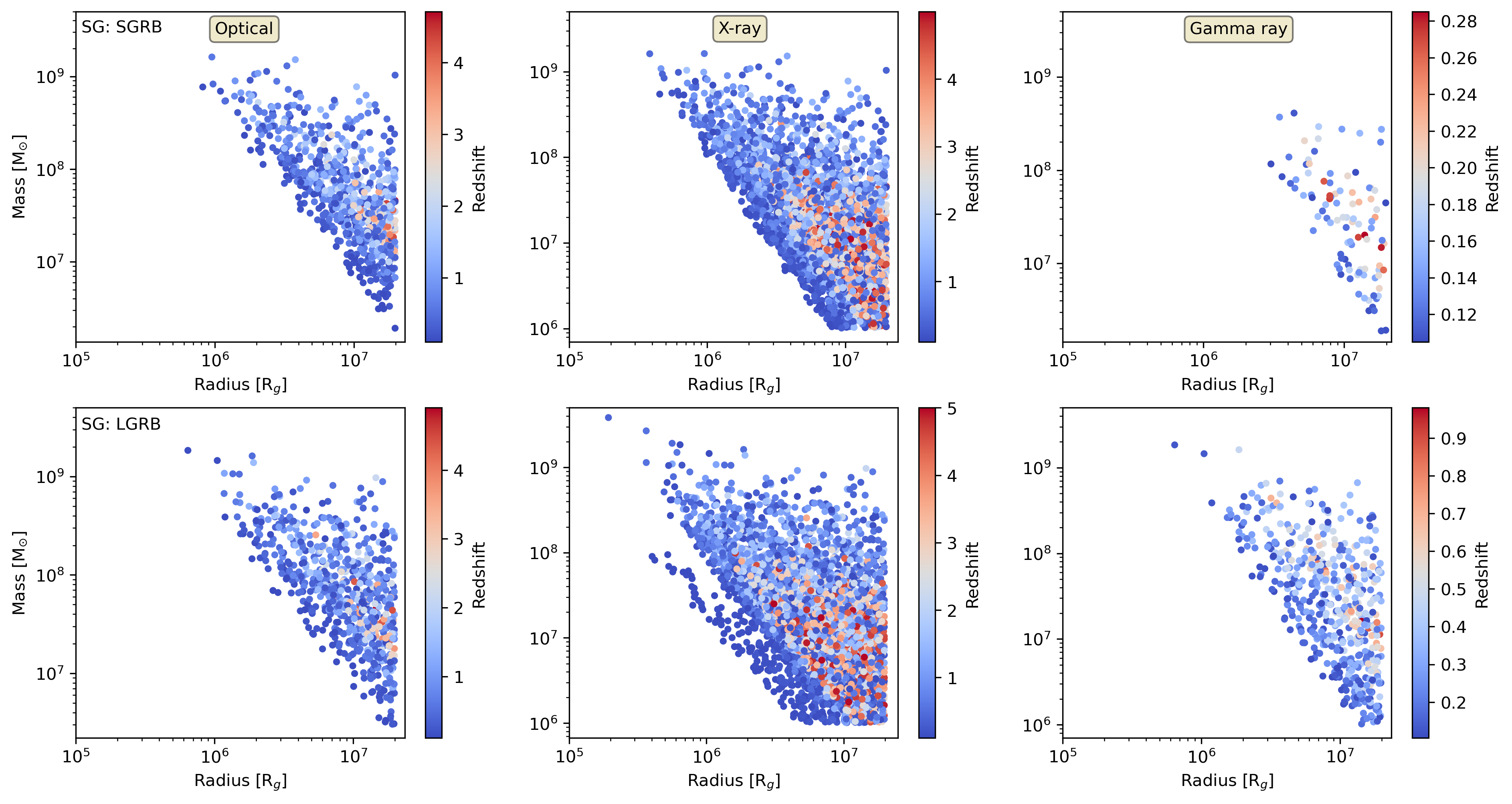}   
    \caption{\small{Afterglow properties of SG model. Scatter plots of mass versus radius for the afterglows of SGRBs (top row) and LGRBs (bottom row) that are above the detection sensitivity threshold in different observational bands: Optical (left), X-ray (middle), and Gamma-ray (right). The color bar represents the redshift of the events, where warmer colors (red) indicate higher redshifts and cooler colors (blue) represent lower redshifts. Most events are detectable at low redshift and from the outer disk regions, with the X-ray band having the highest number of detectable events.}}
    \label{fig:aftglow_detect_SG}
\end{figure*}

\begin{figure*}
    \centering
    \includegraphics[width=1\textwidth]{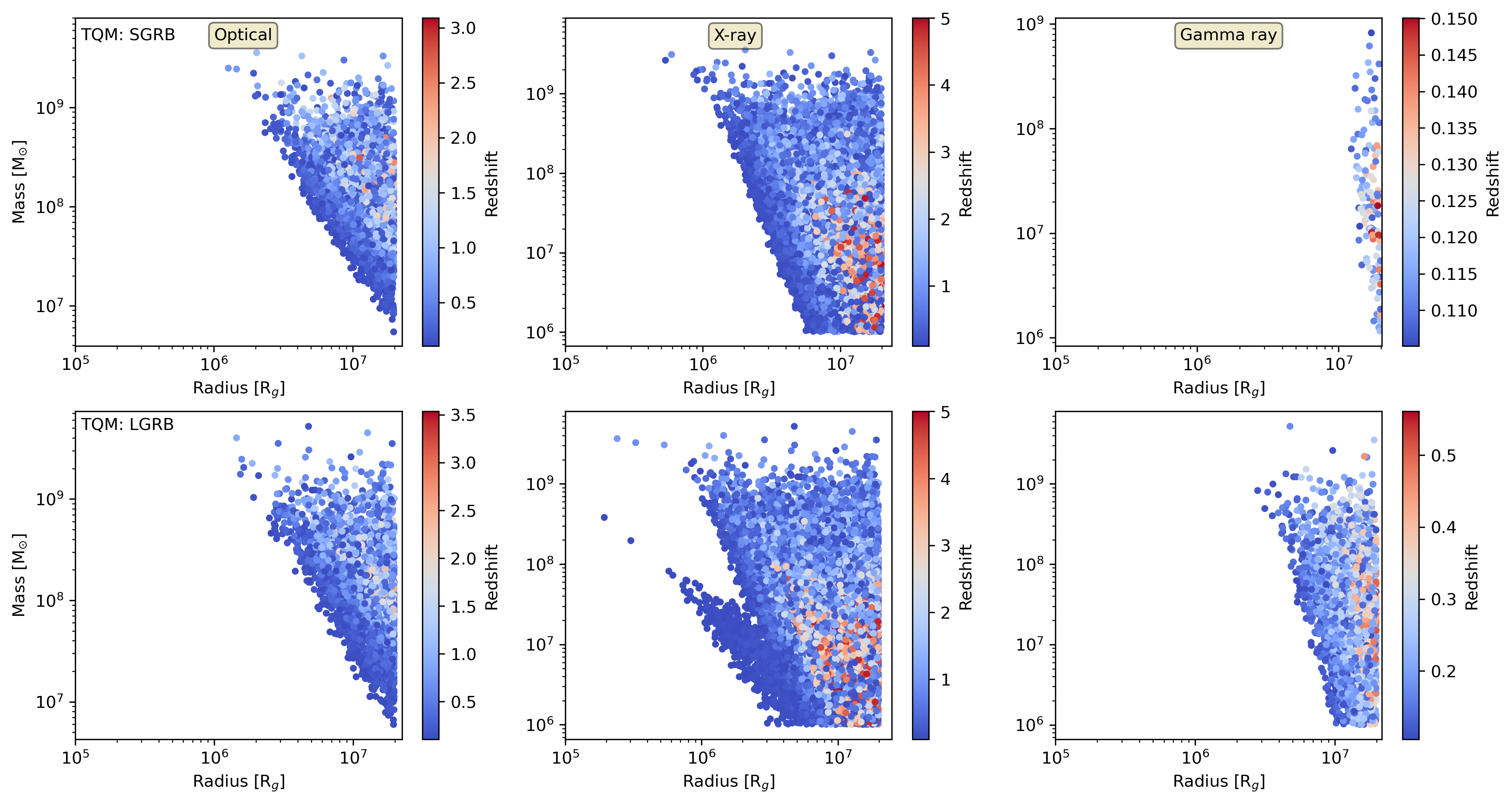}   
    \caption{\small{{Same as Fig.~\ref{fig:aftglow_detect_SG} but for TQM model.}}}
    \label{fig:aftglow_detect_TQM}
\end{figure*}

\begin{figure*}
    \centering
    \includegraphics[width=0.45\textwidth]{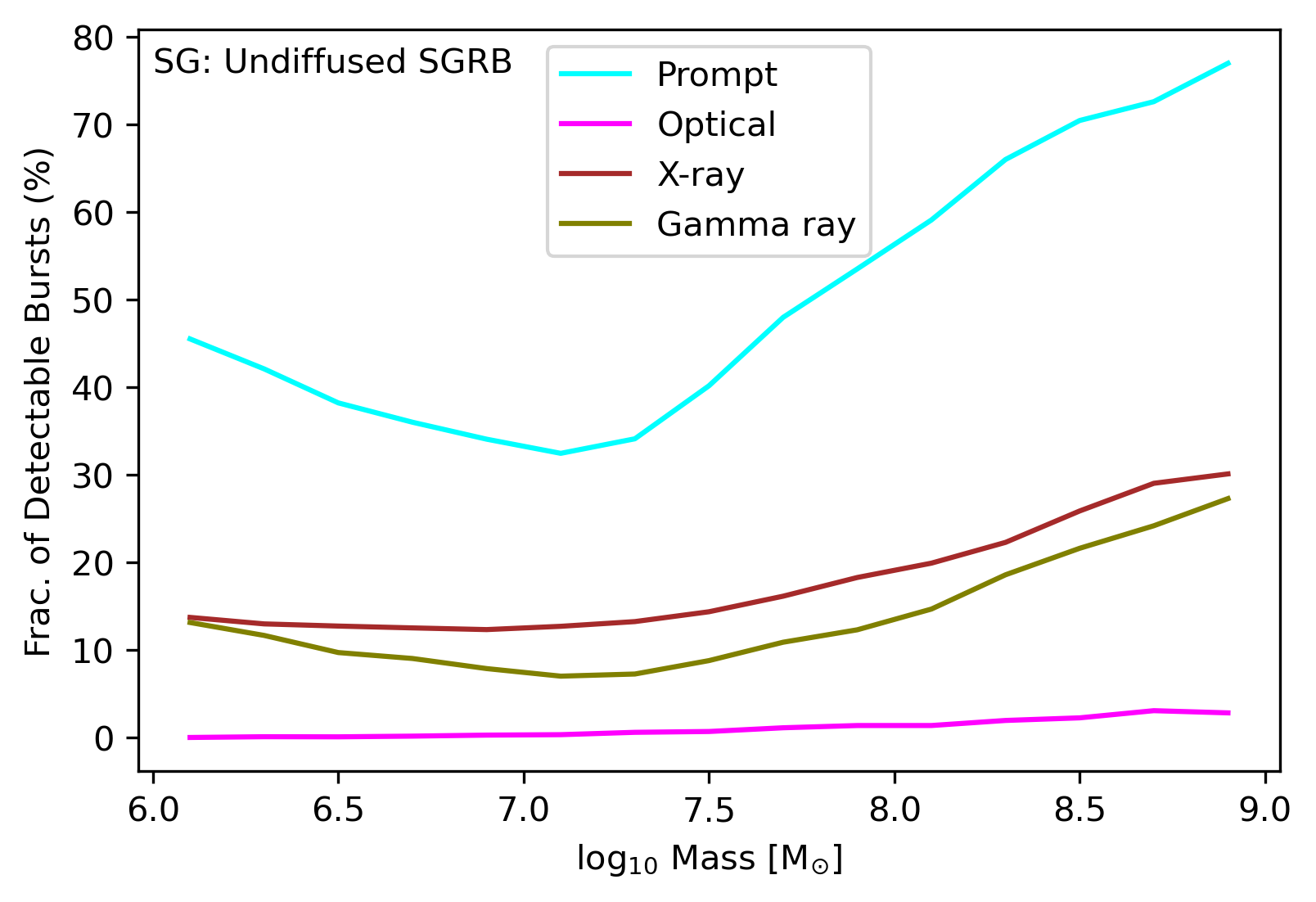}
     \includegraphics[width=0.45\textwidth]{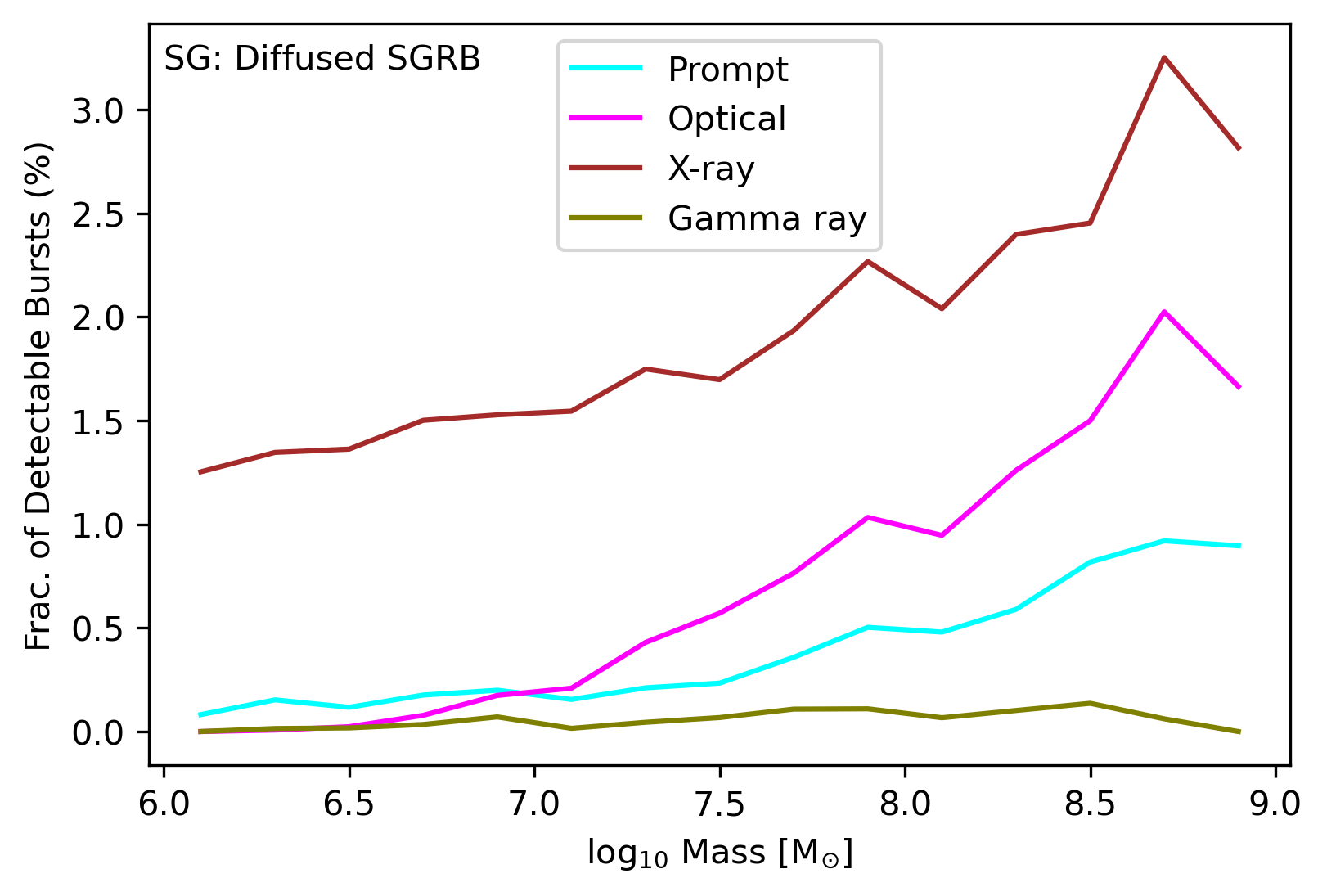}\\     
    \includegraphics[width=0.45\textwidth]{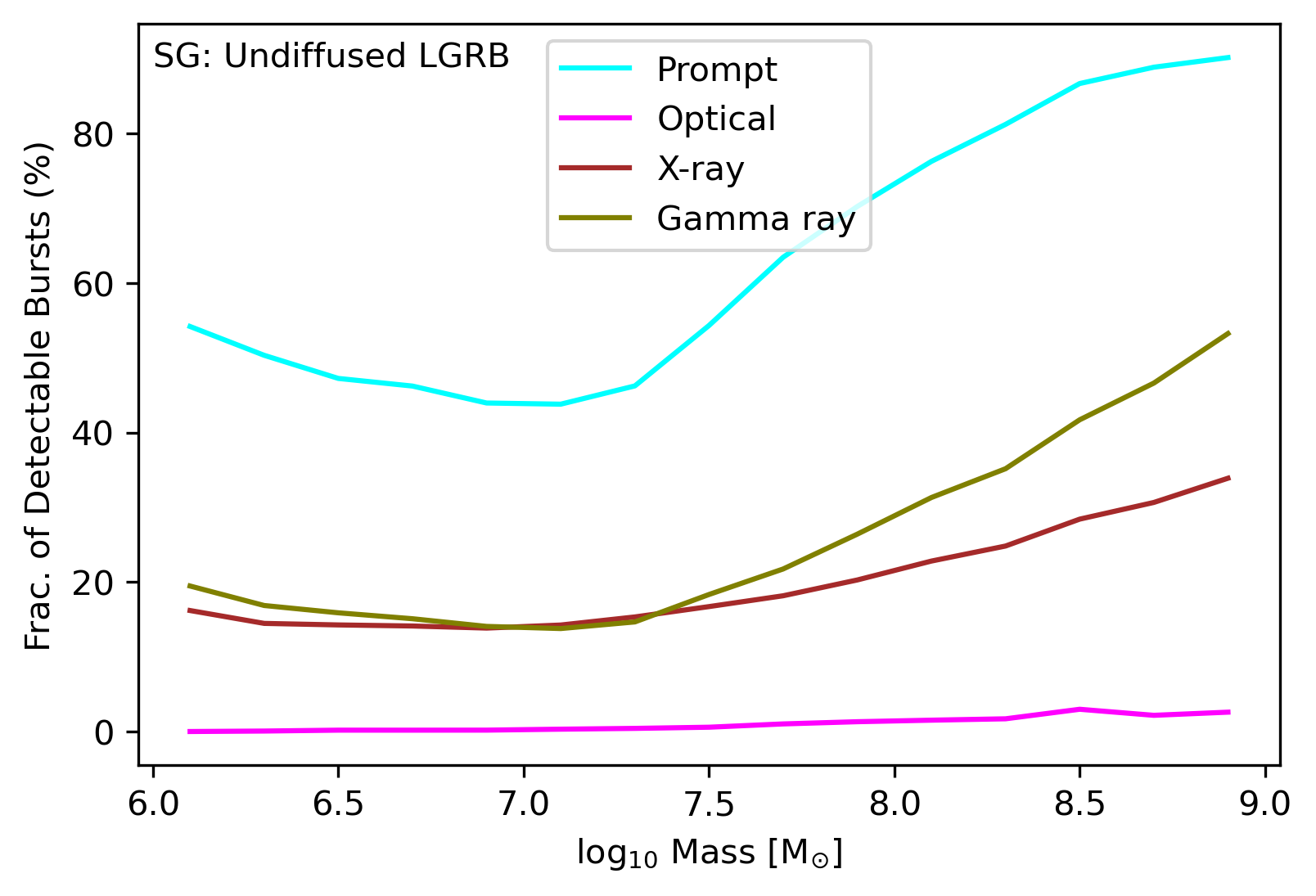}
     \includegraphics[width=0.45\textwidth]{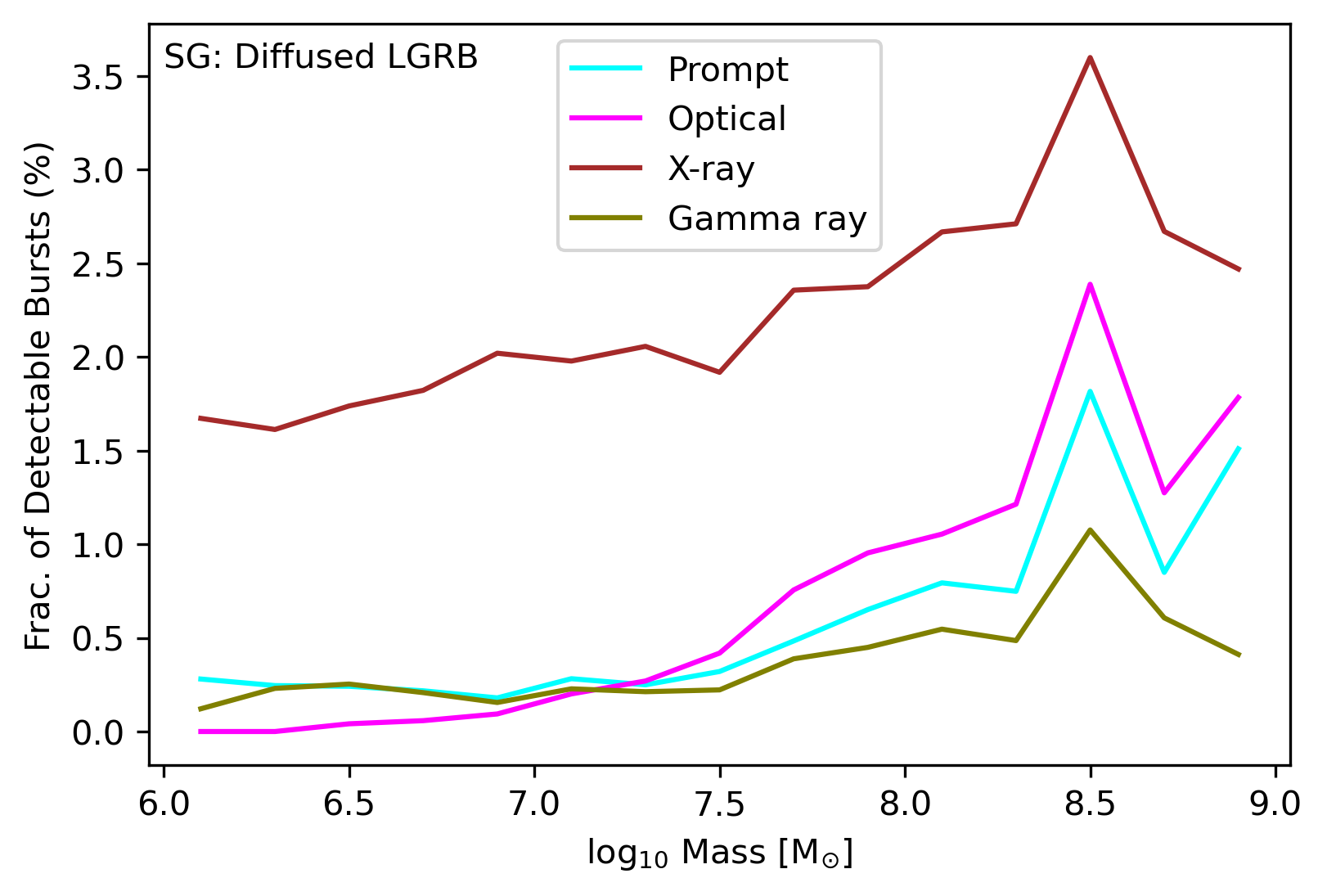}\\      
    \caption{\small{{Fraction of observable events for the SG model as a function of the SMBH mass for Short (top panels) and Long (bottom panels) GRBs, comparing in each case the undiffused model (left panels) with the diffused one (right panels). In the undiffused case, the highest observability across the range of SMBH masses is in the prompt emission. In the diffused case, the overall percentage of detectable events is significantly reduced with some visibility remaining primarily in the X-ray band.}}}
    \label{fig:detect_percent_SG}
\end{figure*}

\begin{figure*}
    \centering
    \includegraphics[width=0.45\textwidth]{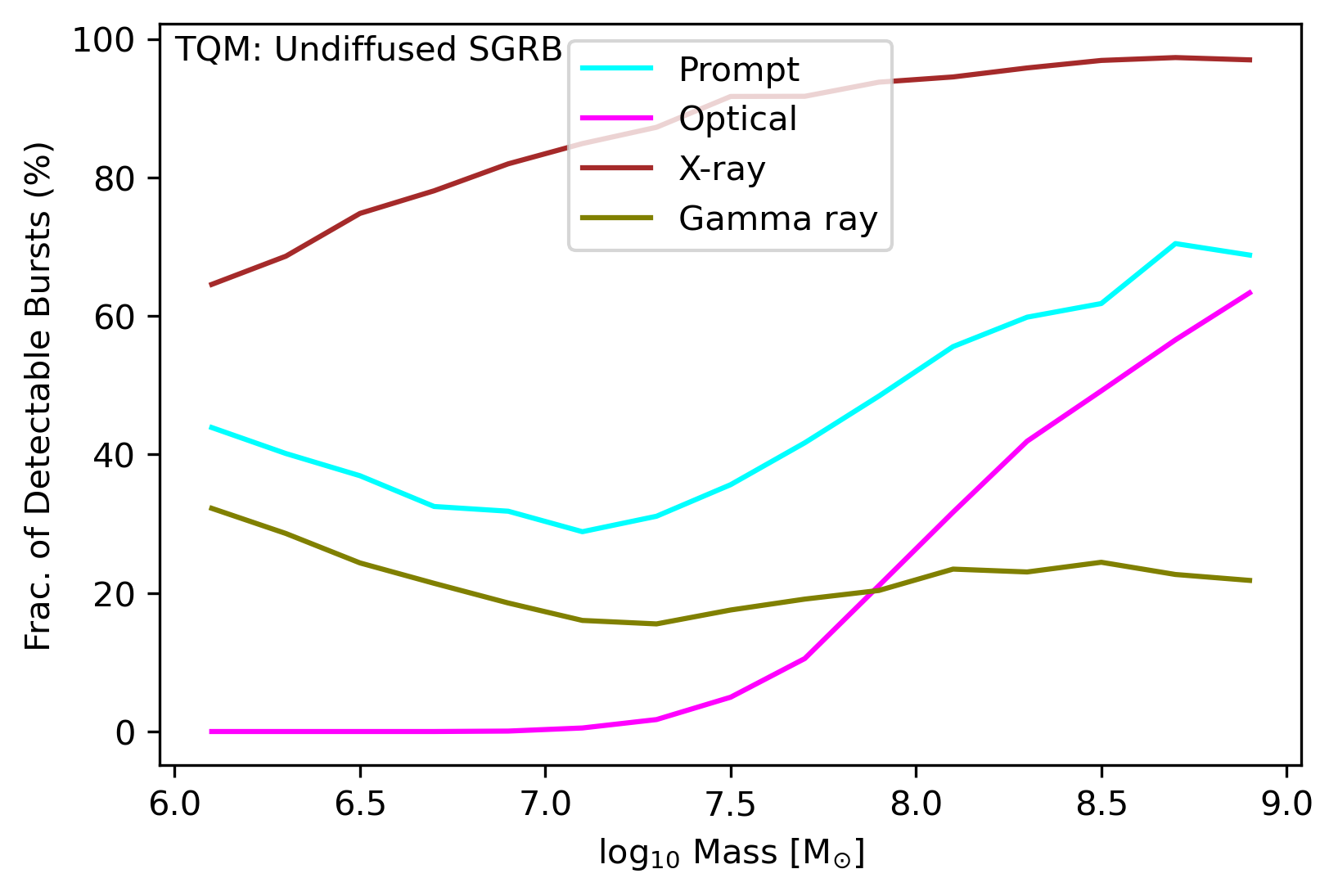}
     \includegraphics[width=0.45\textwidth]{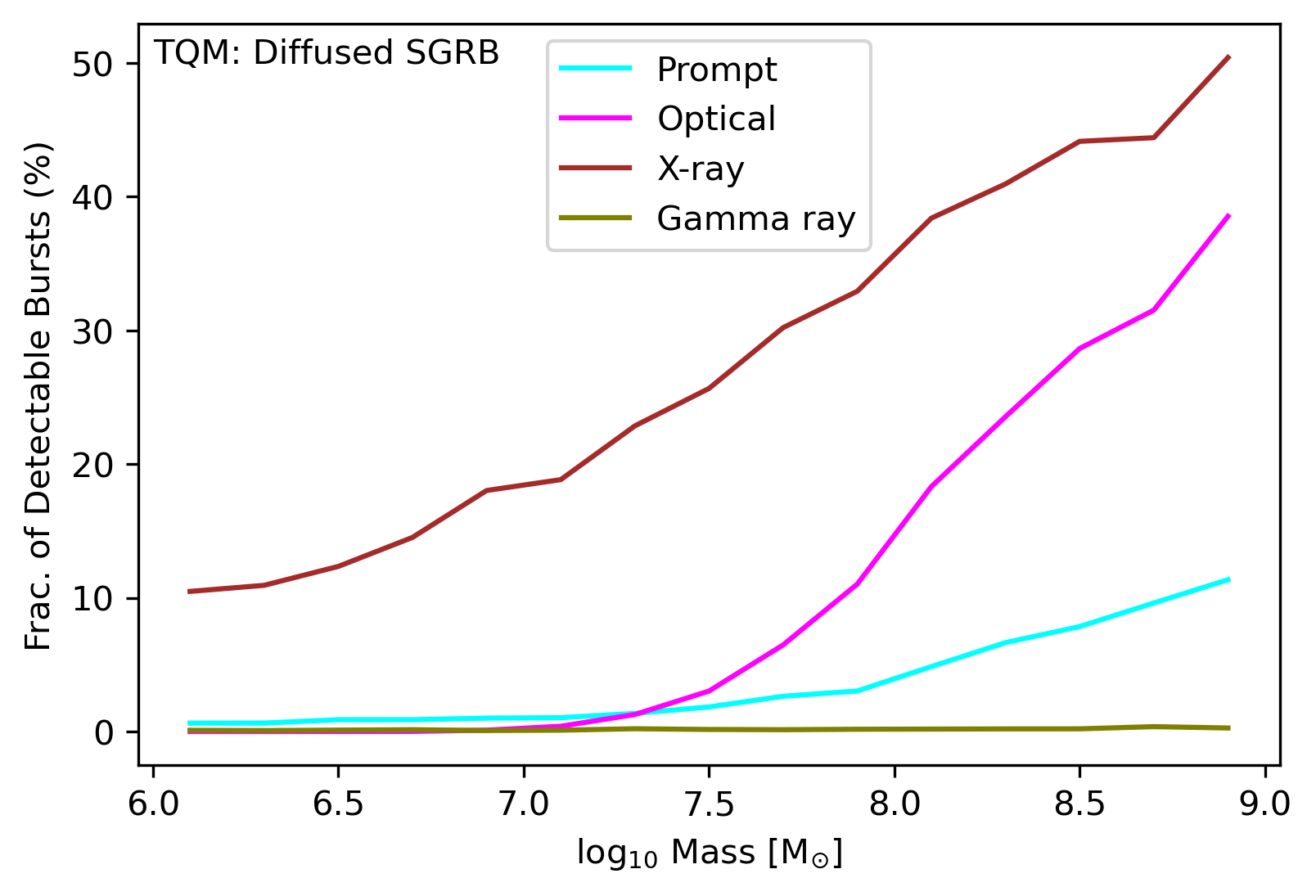}\\     
    \includegraphics[width=0.45\textwidth]{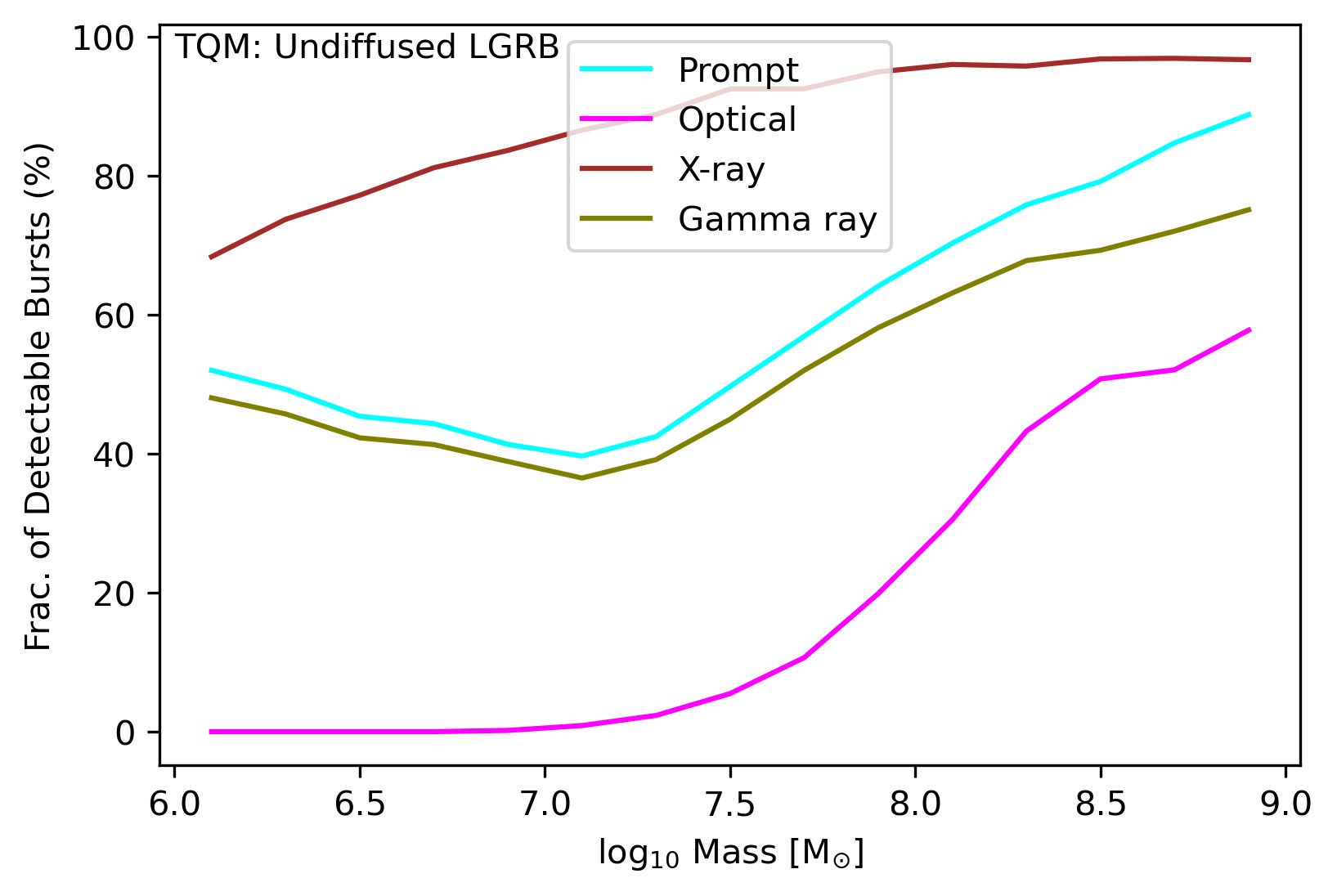}
     \includegraphics[width=0.45\textwidth]{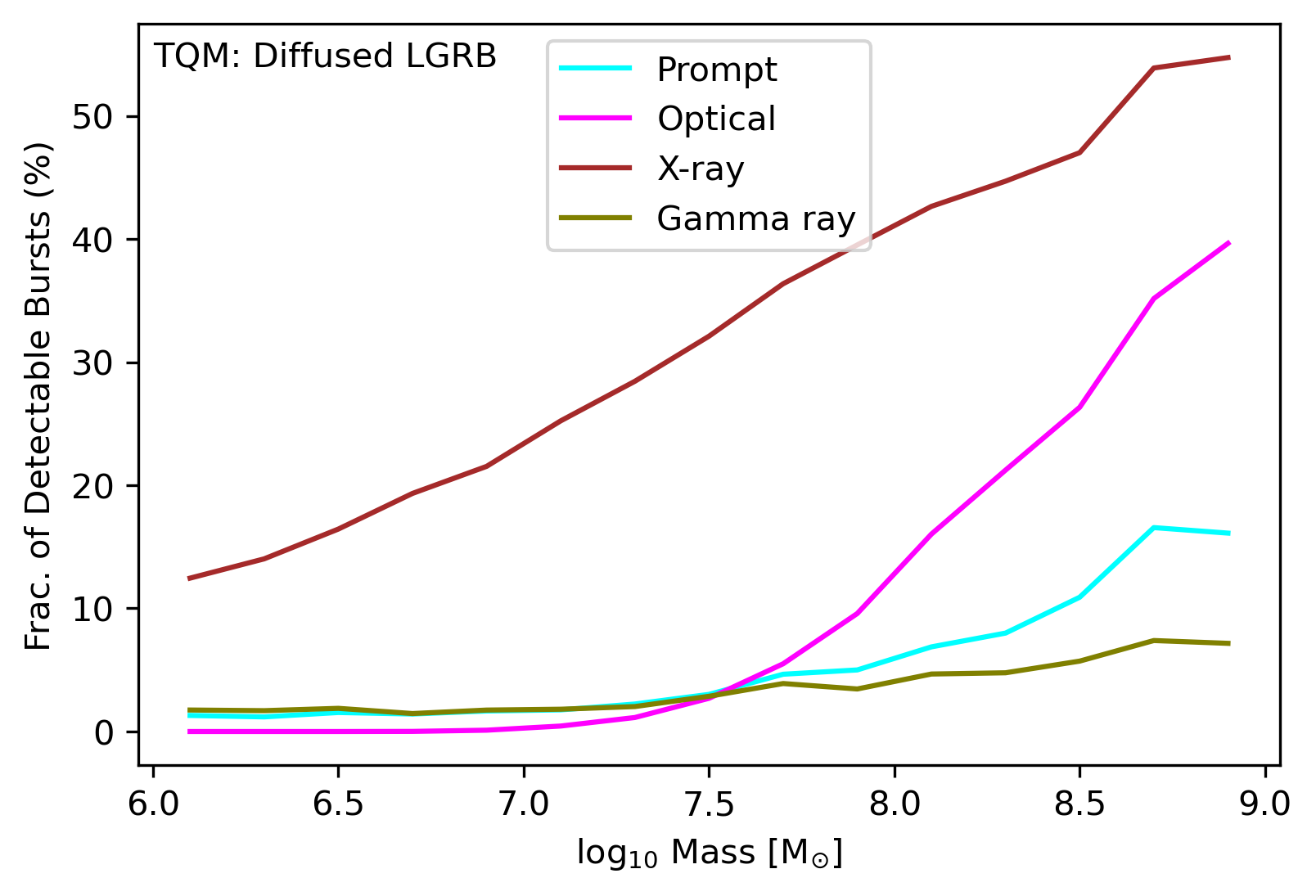}\\      
    \caption{\small{{Same as Fig.~\ref{fig:detect_percent_SG} but for the TQM model. In both the undiffused and diffused scenarios, the highest observability across the range of SMBH masses is in the X-ray band. Detectability in the diffused case is significantly reduced in the lower SMBH mass range, $M_{\rm SMBH} < 10^{7.5} M_{\odot}$.}}}
    \label{fig:detect_percent_TQM}
\end{figure*}

\begin{figure*}
    \centering
    \includegraphics[width=1\textwidth]{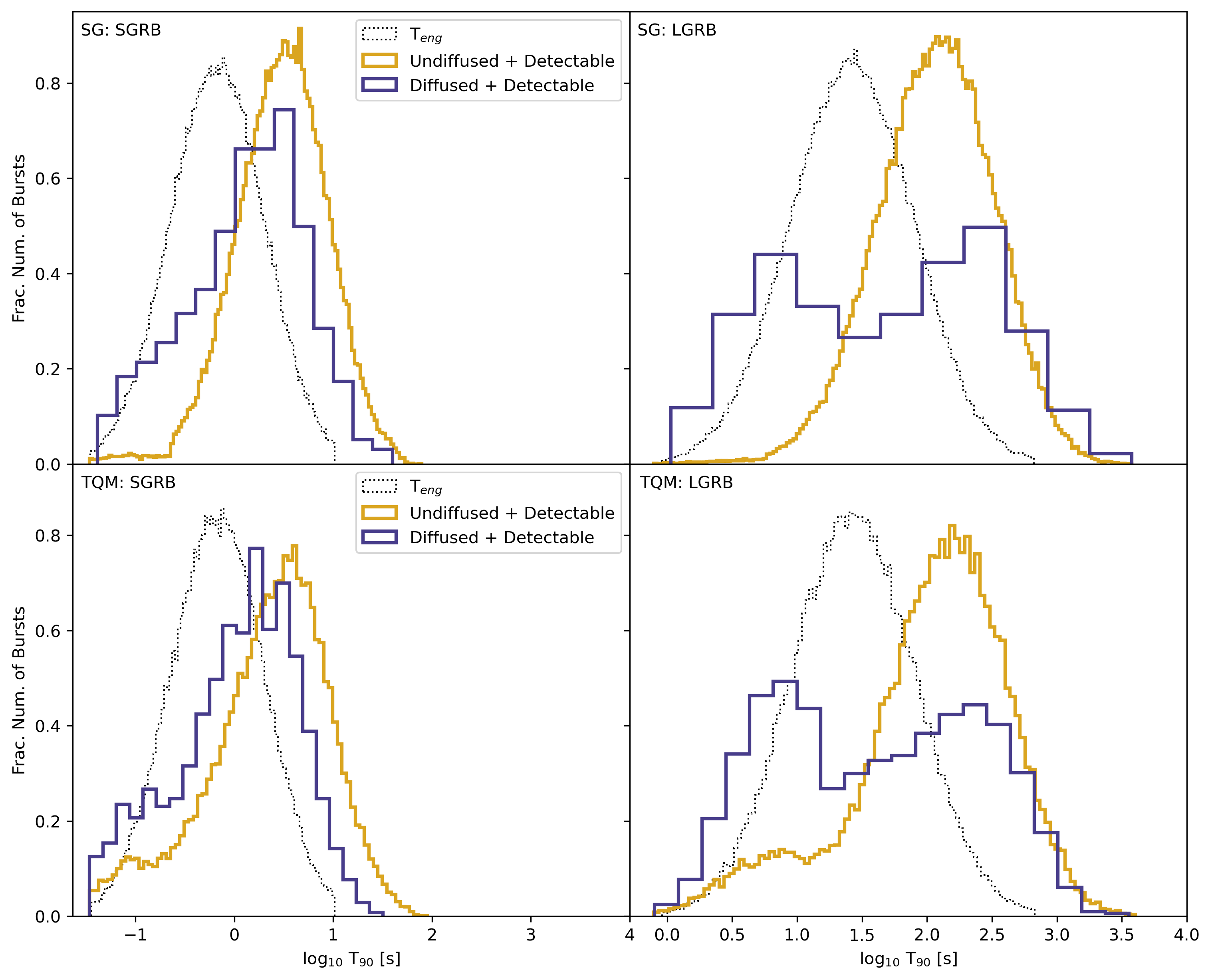}     
    \caption{\small {Probability distributions of 
    the GRB prompt duration $T_{90}$ for sources in an 
    SG disk (top panels) and a TQM disk (bottom panels). Left panels show SGRBs and right ones LGRBs. In all the cases,
     the grey dotted line represents the assumed intrinsic engine duration, while 
     the solid lines show the 
     detectable  $T_{90}$ distributions in the undiffused (navy) and diffused (gold) scenarios.
     }}
    \label{fig:T90}
\end{figure*}
\vspace{1.8in}

{
Given a source of luminosity $L$ at redshift $z$, its bolometric flux at the observer is given by}
\begin{equation}
F = \frac{L}{4\pi D^2_{L}(z)}\,,
\label{eq:flux}
\end{equation}
where $D_L(z)$ is the luminosity distance. 
Correspondingly, the flux density is given by
\begin{equation}
F_{\nu} = \frac{L_{\nu}}{4\pi D^2_{L}(z)(1 + z)}\,,
\label{eq:flux_density}
\end{equation}
where the factor 1 + z accounts for the cosmological effects. The standard cosmological parameters are used: $H_0 = 70$ ~km/s/Mpc, $\Omega_m = 0.3$, and $\Omega_\Lambda = 0.7$.

Fig.~\ref{fig:prmt_ccdf} shows the cumulative probability distribution
of observing the prompt emission from a SGRB (left panel) or a long
GRB (right panel) in an AGN disk, with flux above a certain value. The
threshold corresponding to the sensitivity of the {\em Fermi GBM} detector is
indicated with a vertical black line as a reference. {As expected, the
detection probability is very low in the SG diffused model, hovering well below $\sim 1\%$ for both SGRBs and LGRBs. As for TQM, the detection probability is $2\sim3\%$ for both SGRBs and LGRBs, with the latter having the higher probability. The SG model has a slightly higher fraction of detectability if these transients were to emerge undiffused via a lower density funnel produced by their progenitors. Quantitatively, the SG undiffused model expects a fraction $\sim 41\%$ of SGRBs and of $\sim 53\%$ of LGRBs to be detectable in prompt $\gamma$-rays, while TQM undiffused model expects a fraction of $\sim 38\%$ of SGRBs and of $\sim 49\%$ of LGRBs to be detectable.}      

Fig.~\ref{fig:aftglow_ccdf} correspondingly shows the detection
probabilities from afterglow emission, for four representative bands
and corresponding instrumental sensitivities. {In particular, we use
{\em Fermi GBM}
in $\gamma$ rays, {\em Chandra} with 5 - 6 hours observation time in X-rays, the
{\em Hubble Space Telescope} with ~$\sim$ 1 hour observation time for the optical band,
and {\em VLA} with ~$\sim$ 1 hour observation time for the radio band.}  The detection
probability for the considered sensitivity is practically negligible
in the radio band, even in the most optimistic undiffused model.  As
discussed above, this is due to the heavy suppression due to self
absorption, which is the strongest at longer wavelengths. {The detection probability in the SG model for the diffused case is
marginal in the optical band, just below the percent level, while in the TQM model is around three percent level.} This is because in this band there is still a significant suppression due to self-absorption, which is the same
for the undiffused and the diffused models, since it acts at the source of
emission. The higher energy emission in X-rays and $\gamma$-rays on the
other hand, is significantly less affected by self-absorption of the
emitted radiation, and hence the detection probability varies in a
more pronounced way between the undiffused and the fully diffused scenarios, 
reaching several tens of percent in the former case.

In order to get a better sense of how diffusion influences
observability as a function of the disk radius, SMBH mass, and
redshift, in Fig.~\ref{fig:prmt_detect} we show the locations in the
SMBH mass-radius plane of the simulated GRBs with prompt emission
above the detection threshold. The color of the points, from colder blue
to warmer red, indicate the corresponding redshift.
As expected, most detectable events are produced at the lower redshifts,
$\lesssim 1$, for both the long and the short GRBs. For both groups of
sources, the dominating region of the disk for visibility is in the
outermost part, $\gtrsim 10^6 R_g$, where the density is lower and hence
diffusion is not so extreme. {In the SG model, a fraction of the long GRBs is able to emerge from the innermost region, $R \sim [10^1,10^3]R_g$, where the disk is thin and the fireball has already advanced enough such that the external shock radius exceeds the scale height, $R_{em}/H > 1$. The same condition is possible in the TQM model around the $R = 10^4 R_g$ region, but is almost negligible. Note that this condition is even more unlikely for short GRBs, as their isotropic equivalent energy, $E_{\rm iso}$, is two orders of magnitude lower than that of long GRBs.}

Similarly, Figures~\ref{fig:aftglow_detect_SG} and~\ref{fig:aftglow_detect_TQM} show the locations of the detectable transients for the afterglow emission (for the same
detection thresholds used in Fig.~\ref{fig:aftglow_ccdf})
where radio has been omitted since it is entirely below detection.
Not surprisingly, the outer regions of the disk are the most populated ones.
Similar to the prompt emission, detectability is much more
significant at high SMBH masses. Note that in X-rays there is also
a non-negligible contribution from relatively higher redshifts.

To better visualize the dependence of observability on SMBH mass,
in Figures~\ref{fig:detect_percent_SG} and~\ref{fig:detect_percent_TQM} we show the fraction of detectable
GRBs as a function of SMBH mass. {In both the undiffused and diffused scenarios, AGNs with larger SMBH masses dominate the observable GRB fraction. This is an immediate corollary of the assumption of the number of GRB transients being proportional to the disk mass. In the absence of line-of-sight opacity, the number density is the driving factor in determining the emission.  The higher number density increases the mass of material encountered by the relativistic jet, enhancing energy dissipation through stronger external shocks. This leads to amplified magnetic fields, which boost synchrotron radiation and ultimately raise the peak luminosity. However, self-absorption becomes increasingly more important in denser media, especially at lower frequencies. Considering the number density distributions in the SG and TQM models (SG peak $\sim 10^{15} \rm cm^{-3}$ and TQM peak $\sim 10^8 \rm cm^{-3}$), it is not surprising to see that the highest observability in the SG model occurs in the prompt emission. While the prompt emission remains highly observable in the TQM model, the X-ray band exhibits the highest detectability, with a significant probability in the optical band as well. In the diffused case, detectability is diminished in both the SG and TQM models as the optical depth suppresses the emission. The general distribution of optical depth in the SG model, which is several orders of magnitude higher compared to the TQM model, results in a significantly more substantial reduction in detectability for SG.}

For the GRBs which are able to emerge, it is interesting to
investigate the distribution of the duration times
$T_{90}$. Reminding the discussion earlier, \citet{Lazzati2022}
identified a critical density (whose exact value depends on the burst
energetic and Lorentz factor), above which the external shock forms
before the internal one, resulting in a slightly dimmer GRB but with
longer duration $T_{90}$.  The statistical outcome of the duration
distribution is shown in Fig.~\ref{fig:T90}, for both the short GRBs
(left panel) and the long ones (right panel). In both cases, the
intrinsic distribution of engine durations is compared to that of
$T_{90}$. For both the undiffused and the diffused scenarios, we display
both the intrinsic distribution as well as the distributions
corresponding to the subset of GRBs whose flux is above the detection
limit of Fermi. For LGRBs, the peak of the distribution shifts from several tens of 
seconds to several hundreds of seconds. {These could hence be mis-classified
as belonging to the class of very long GRBs \citep{Gendre2013}.}
Conversely, short GRBs, whose intrinsic distribution peaks at a fraction of a
second, are found to peak in $T_{90}$ at several seconds, hence making
these short GRBs rather look as long GRBs, as argued for the specific case
of GRB191019A (\citealt{Lazzati2023}, but see \citealt{Stratta2024}).
Further confirmation for an AGN-disk origin can be provided by longer wavelength observations aimed at detecting the afterglow: counterparts in X-rays are expected  if there was a $\gamma$-ray detection. Optical counterparts may be observed with $\sim 40-60\%$  probability only in the most massive disks (due to the lower density), while a radio counterpart is extremely unlikely.

For the diffused distributions, shown with navy lines in
Fig.~\ref{fig:T90}, the outcome is slightly more complex, since
'stretching' of the effective burst duration is produced not only at
the source (similarly to the undiffused case), but also as a result of
diffusion. The majority of the diffused sources would be emerging on
very long timescales, on the order of several years, and similar for
both the long and the short GRBs, since, for these, the diffusion timescales
largely dominates over the intrinsic duration. These very long GRBs
are correspondingly very dim and therefore unobservable with current
telescopes. The small subset of detectable bursts in the diffusive scenario
is the one encountering smaller optical depths, and thus with a diffusive
timescale which is generally smaller than the $T_{90}$ duration.
The $T_{90}$ distribution of observable bursts in the diffusive scenario
appears structured, with a tail/smaller second peak tracking $T_{\rm eng}$
from the outermost regions of the disk with both low optical depth and
density below the critical one to significantly affect the duration of
the intrinsic emission. {In the LGRB case, the diffused $T_{90}$ distribution is broader than that of SGRBs because LGRBs have inherently longer intrinsic engine durations, allowing for a wider range of diffusion timescales to influence $T_{90}$. The diffusion process, governed by the optical depth and number density, introduces additional variability in $T_{90}$.}
We remind that, for better visualization, all the distributions in
Fig.~\ref{fig:T90} have been normalized to 1. However, the normalization
factor of the diffused and detectable GRBs is much smaller than for the other
distributions, as shown in Figures~\ref{fig:detect_percent_SG} and~\ref{fig:detect_percent_TQM}.

\section{Summary and Discussion}
\label{sec:concl}

{In this work we have studied the population and detectability of GRBs emerging from AGN disks. To characterize the disk properties, we have adopted two widely used models for AGN disks:  SG and TQM. For both models, we have considered two extreme cases for the propagation of the radiation: completely undiffused and fully diffused, in order to straddle the realistic situations which are likely to be in between, but which are not easy to model from first principles, since conditions can be highly non-uniform across the whole cosmological AGN population. In both cases, the intrinsic
prompt and afterglow emission is calculated based on local disk properties. }

In the diffused case, GRBs emerging from outer disk regions, where density and  optical depth are relatively lower, are more likely to be detectable, whereas those originating in the inner regions face significant suppression. Specifically, we expect to observe GRBs mostly from low redshifts and AGN disks with large SMBH mass,  $\gtrsim 10^{7.5} M_{\odot}$ and in outer regions of the disk, $R \sim [10^6,10^7] R_g$. Similarly, in the undiffused case, we expect to observe GRBs mostly from low redshifts and AGN disks with large SMBH masses. However, there is also a substantial contribution from intermediate regions of the disk. 

For both disk models, GRBs would most likely be detectable via the prompt $\gamma$-ray emission and the afterglow emission in the X-ray band. The SG model, characterized by regions of higher densities and optical depths, results in suppressed detectability, particularly in the diffused scenario. However, in the undiffused case, the prompt emission dominates as the primary observational window due to the stronger external shocks that enhance synchrotron radiation. On the other hand, the TQM model, with its orders of magnitude lower densities than the SG model, and less extreme optical depths, favors detectability across a wider range of wavelengths, especially in the X-ray band and with some contribution in the optical waveband.  These results underscore the crucial dependence of GRB detectability on disk models, where SG disk environments amplify prompt emission through higher densities, whereas TQM environments provide a broader wavelength range of detectable signals due to their relatively lower optical depths. The differences between the two models are so consequential that it might be possible to use even a few GRB detections to learn about the AGN disk structure. 
Our results are in broad agreement with those of \citet{Zhang2024b} in that observability is favored for more energetic GRBs, and that there is a strong dependence on the disk properties. However, the specific details vary since they explored the hydrodynamic properties of the jets in dense environments, while we focused on the emission properties and line-of-sight scattering opacity.

The $T_{90}$
distributions of both long and short GRBs peak at longer times compared to their low-density, galactic counterparts. Observationally, this can result in intrinsically short GRBs to be mis-classified as long, and in typically long GRBs to be mis-classified as very long.

All our results for the observable quantities are provided in terms of probability distributions. This allows us to factor out the dependence on the absolute number of stars and compact objects in AGN disks, which are uncertain quantities.
In turn, if AGN observations are able to calibrate the number of GRBs from their disks, then our probabilities would allow to place constraints on the numbers of stars and compact objects in them.

In addition to GRBs, AGN disks are expected to host a variety of electromagnetic transients that could produce overlapping observational signatures. These include tidal disruption events from the central SMBH \citep{Metzger2012},
micro tidal disruption events from stellar-mass BHs \citep{Yang2022}, 
jets from merging binary BHs \citep{Tagawa2023m} as well as isolated BHs undergoing  hyperaccretion \citep{Pan2021,Tagawa2023s,Liu2024}, 
supernovae \citep{Grishin2021},
interacting kilonovae \citep{Ren2022},
and accretion-induced collapse of neutron stars  \citep{Perna2021b,Zhang2024} and white dwarves \citep{Zhu2021WD}. Each of these events could contribute to the overall transient activity observed in AGN environments, and distinguishing between them will require careful analysis of their light curves, spectral signatures, and temporal evolution. As more data becomes available, we expect to refine the criteria for identifying these AGN-hosted GRBs and disentangling them from the broader transient population.

\section*{Acknowledgments}

{RP gratefully acknowledges support by NSF award AST-2006839.}

\bibliographystyle{aasjournal}

\bibliography{biblio}



\end{document}